\documentclass[fleqn,10pt]{wlscirep}

\usepackage[utf8]{inputenc}
\usepackage[T1]{fontenc}
\usepackage{graphics}
\usepackage{bm}
\usepackage{xcolor}
\usepackage{todonotes}
\usepackage{hyperref}
\usepackage{xcolor}
\usepackage{lineno}
\usepackage{setspace} 
\usepackage{multibib}
\usepackage[flushleft]{threeparttable}

\hypersetup{final,colorlinks=true,linkcolor=blue,citecolor=blue,urlcolor=blue}

\newcites{S}{Methods references} 

\graphicspath{{pic/}}

\newcommand{\tento}[1]{10$^{#1}$}

\newcommand{\mnras}{Monthly Notices of the Royal Astronomical Society}
\newcommand{\aj}{Astronomical Journal}
\newcommand{\apj}{Astrophysical Journal}
\newcommand{\apjl}{Astrophysical Journal Letters}
\newcommand{\aap}{Astronomy \& Astrophysics}

\title{A dusty veil shading Betelgeuse during its Great Dimming}

\author[1,2,*]{M.~Montargès}
\author[2]{E.~Cannon}
\author[3]{E.~Lagadec}
\author[4, 2]{A.~de Koter}
\author[1]{P.~Kervella}
\author[5, 6]{J.~Sanchez-Bermudez}
\author[7]{C.~Paladini}
\author[5]{F.~Cantalloube}
\author[2, 8]{L.~Decin}
\author[7]{P.~Scicluna}
\author[9]{K.~Kravchenko}
\author[10]{A.~K.~Dupree}
\author[11]{S.~Ridgway}
\author[12]{M.~Wittkowski}
\author[13, 14]{N.~Anugu}
\author[15]{R.~Norris}
\author[16, 17]{G.~Rau}
\author[1]{G.~Perrin}
\author[3]{A.~Chiavassa}
\author[14]{S.~Kraus}
\author[18]{J.~D.~Monnier}
\author[3]{F.~Millour}
\author[19, 18]{J.-B.~Le Bouquin}
\author[7]{X.~Haubois}
\author[3]{B.~Lopez}
\author[3]{P.~Stee}
\author[16]{W.~Danchi}

\affil[1]{LESIA, Observatoire de Paris, Université PSL, CNRS, Sorbonne Université, Université de Paris, 5 place Jules Janssen, 92195 Meudon, France}
\affil[2]{Institute of Astronomy, KU Leuven, Celestijnenlaan 200D B2401, 3001 Leuven, Belgium}
\affil[3]{Université Côte d'Azur, Observatoire de la Côte d'Azur, CNRS, Laboratoire Lagrange, Bd de l'Observatoire, CS 34229, 06304 Nice cedex 4, France}
\affil[4]{Anton Pannenkoek Institute for Astronomy, University of Amsterdam, 1090 GE, Amsterdam, The Netherlands}
\affil[5]{Max Planck Institute for Astronomy, Königstuhl 17, 69117, Heidelberg, Germany}
\affil[6]{Instituto de Astronom\'ia, Universidad Nacional Aut\'onoma de M\'exico, Apdo. Postal 70264, Ciudad de M\'exico, 04510, M\'exico}
\affil[7]{European Southern Observatory, Alonso de Cordova 3107, Vitacura, Santiago, Chile}
\affil[8]{School of Chemistry, University of Leeds, Leeds LS2 9JT, UK}
\affil[9]{Max Planck Institute for extraterrestrial Physics, Giessenbachstraße 1, D-85748 Garching, Germany}
\affil[10]{Center for Astrophysics, Harvard \& Smithsonian, 60 Garden Street, Cambridge, MA 02138, USA}
\affil[11]{NSF’s National Optical-Infrared Astronomy Research Laboratory, PO Box 26732, Tucson, AZ 85726-6732, USA}
\affil[12]{European Southern Observatory, Karl-Schwarzschild-Str. 2, 85748, Garching bei München, Germany}
\affil[13]{Steward Observatory, 933 N. Cherry Avenue, University of Arizona, Tucson, AZ, 85721, USA}
\affil[14]{University of Exeter, School of Physics and Astronomy, Stocker Road, Exeter, EX4 4QL, UK}
\affil[15]{Physics Department, New Mexico Institute of Mining and Technology, 801 Leroy Place, Socorro, NM 87801, USA}
\affil[16]{NASA Goddard Space Flight Center, Exoplanets \& Stellar Astrophysics Laboratory, Code 667, Greenbelt, MD 20771, USA}
\affil[17]{Department of Physics, Catholic University of America, Washington, DC 20064, USA}
\affil[18]{Department of Astronomy, University of Michigan, Ann Arbor, MI, 48109, USA}
\affil[19]{Univ. Grenoble Alpes, CNRS, IPAG, 38000 Grenoble, France}

\affil[*]{e-mail: miguel.montarges@observatoiredeparis.psl.eu}

\begin{abstract}
	
	Red supergiants represent the most common final stage of the evolution of stars with initial masses between 8 and 30-35 times the mass of the Sun\cite{2012A&A...537A.146E}. During this phase of lifetime lasting $\approx$ 10$^{5}$\,yrs\cite{2012A&A...537A.146E}, they experience substantial mass loss of unknown mechanism\cite{2015A&A...575A..50A}. This mass loss can affect their evolutionary path, collapse, future supernova light curve\cite{2018MNRAS.476.2840M}, and ultimate fate as a neutron star or a black hole\cite{2015A&A...575A..60M}. 
	From November 2019 to March 2020, the second closest red supergiant (RSG, $222^{+48}_{-34}$\,pc\cite{2017AJ....154...11H,2020ApJ...902...63J}) Betelgeuse experienced a historic dimming of its visible brightness, witnessed worldwide. Usually between 0.1 and 1.0 mag, it went down to $1.614 \pm 0.008$ mag around 7-13 February 2020 \cite{2020ATel13512....1G}. 
	Here we report high angular resolution observations showing that the southern hemisphere of the star was ten times darker than usual in the visible. Observations and modeling support the scenario of a dust clump recently formed in the vicinity of the star due to a local temperature decrease in a cool patch appearing on the photosphere. The directly imaged brightness variations of Betelgeuse evolved on a timescale of weeks. 
	This event suggests that an inhomogeneous component of red supergiant mass loss\cite{2011A&A...531A.117K} is linked to a very contrasted and rapidly changing photosphere.
	
\end{abstract}

\begin{document}
	
	\flushbottom
	\maketitle
	
	\thispagestyle{empty}
	
	\noindent Received 6 November 2020; accepted 13 April 2021
	
	
	\section*{Main paper}
	
\begin{figure}[ht!]
	\centering
	\includegraphics[width=\columnwidth]{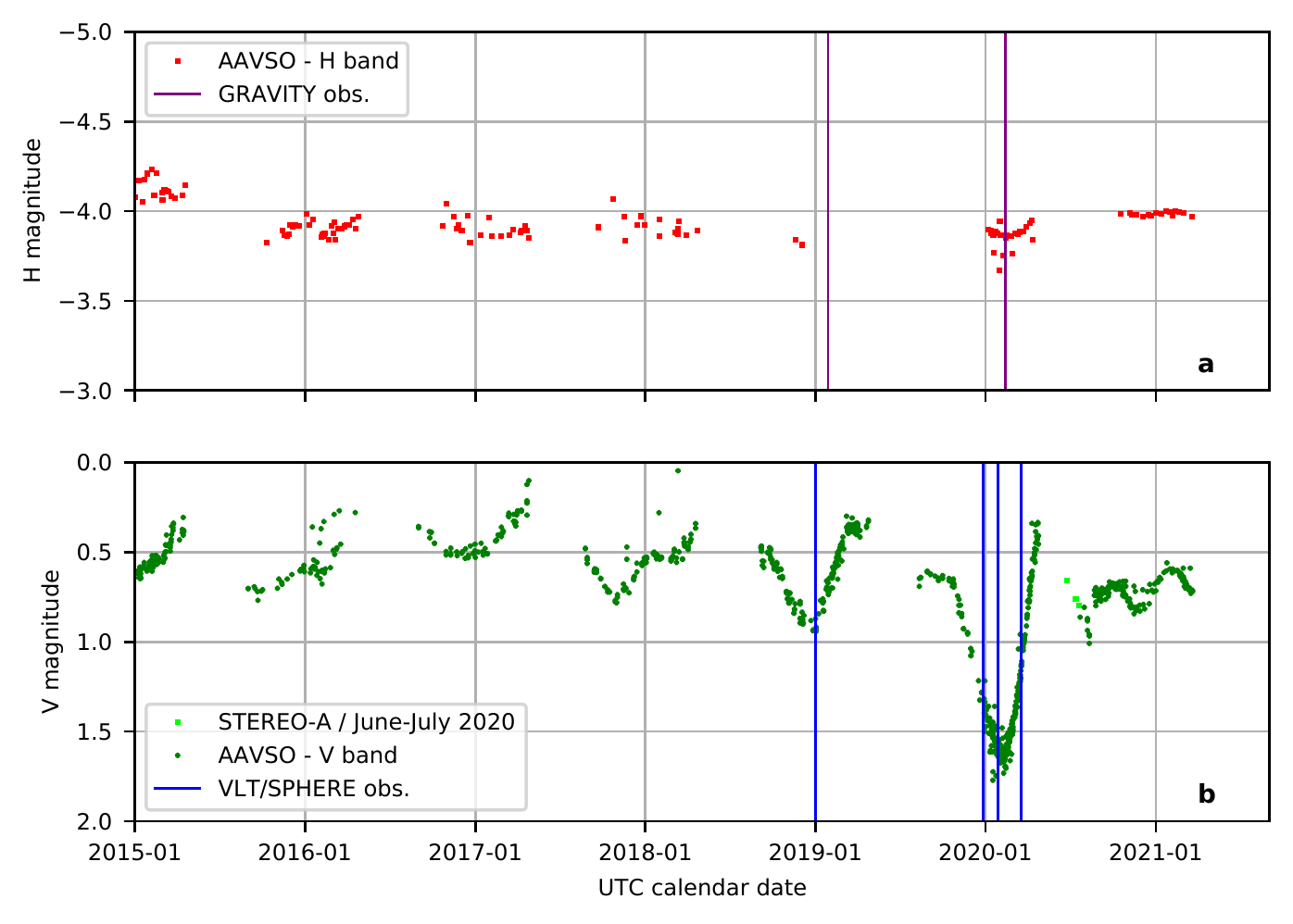}
	\caption{\textbf{Light curve of Betegeuse over the past six years.} Data are obtained from the AAVSO and STEREO-A\cite{2020ATel13901....1D}. The dates of the observations presented in this article are indicated by the vertical lines. Panel \textbf{a} corresponds to the near-infrared, panel \textbf{b} to the visible.}
	\label{Fig:AAVSO_curve}
\end{figure}

	
	We obtained direct high spatial resolution observations using VLTI/GRAVITY and VLT/SPHERE. Details on the data aquisition and reduction are available in the \textit{Methods} section. The SPHERE-ZIMPOL images provide the only resolved reconnaissance of the stellar disk and its nearby surroundings a year before and through this Great Dimming event. Observations were secured before (January 2019) and during the dimming (December 2019, January and March 2020, see also Fig.~\ref{Fig:AAVSO_curve}).\\
	
\begin{figure}[ht!]
	\centering
	\includegraphics[width=\columnwidth]{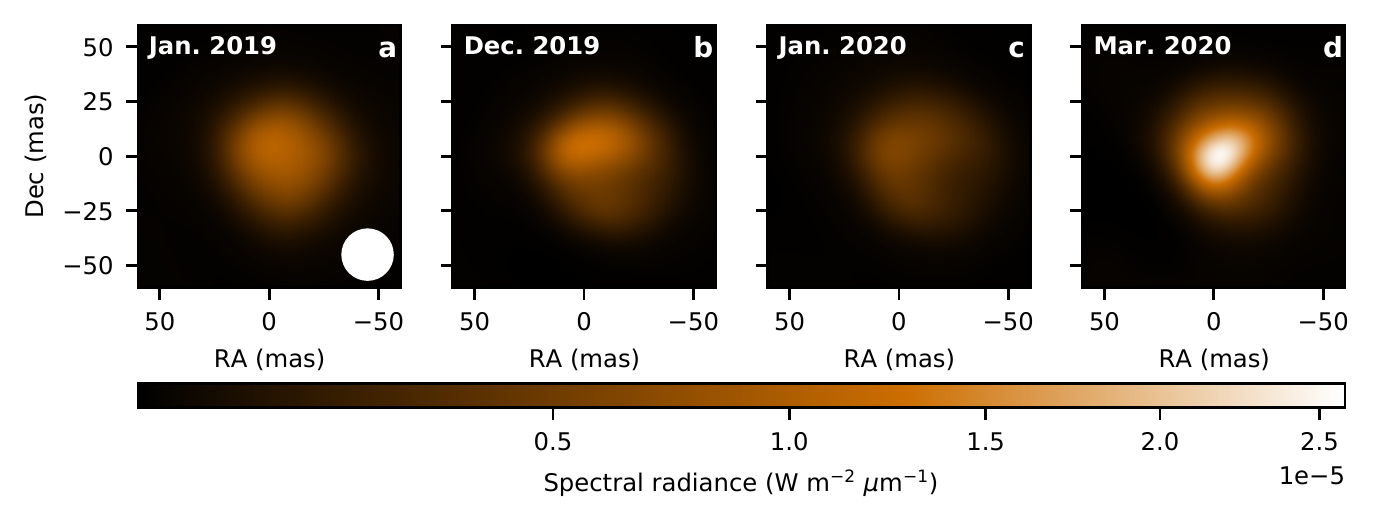}
	\caption{\textbf{VLT/SPHERE-ZIMPOL observations of Betelgeuse after deconvolution in the Cnt\_H$\alpha$ filter.} North is up and East is left. The beam size of ZIMPOL is indicated by the white disk at the bottom right corner of image \textbf{a}. We used a power-law scale intensity with index 0.65 to enhance the contrast.}
	\label{Fig:ZIMPOL_SPHERE}
\end{figure}

	
	\noindent The deconvolved ZIMPOL images of Betelgeuse for the four epochs are presented in Fig.~\ref{Fig:ZIMPOL_SPHERE}. The photosphere of Betelgeuse is clearly resolved in each image, with deviations from circular symmetry. Earlier imaging in March 2015, published in 2016\cite{2016A&A...585A..28K}, showed an elongation along the North-East to South-West axis. The same shape is apparent in our January 2019 image (Fig.~\ref{Fig:ZIMPOL_SPHERE}a) but less pronounced. In both December 2019 and January 2020 (Fig.~\ref{Fig:ZIMPOL_SPHERE} b and c), in all filters, the Southern hemisphere of the star has a peak brightness over 10 times less than the Northern hemisphere.\\
	
	
	\noindent Four scenarios could explain the Great Dimming of Betelgeuse: (1) a (potentially local) decrease of the effective temperature of the star, making it fainter; an occultation by dust (2) either newly formed or (3) transiting in front of the star, or (4) a change of the angular diameter. Each scenario is checked against our observations.
	
	\noindent A transiting dusty clump scenario is rejected as such a clump should change quadrant before and after the deepest dimming, while our observations in January and March 2020 show the dark area remained in the South-Western quadrant. A change in diameter is ruled out by our VLTI/GRAVITY and VLT/SPHERE-IRDIS observations. We measured uniform disk (UD) angular diameters of $\theta_\mathrm{UD} = 42.61 \pm 0.05$\,mas in January 2019, and $\theta_\mathrm{UD} = 42.11 \pm 0.05$\,mas in February 2020, well within the range explored during the past 30 years\cite{2011A&A...529A.163O}, and far from the 30\% variation that would have been required to reproduce the visible dimming.
	
	\noindent Dust has been inferred from spectrophotometric observations\cite{2020ApJ...891L..37L}. However, later sub-millimeter observations and TiO photometry\cite{2020ApJ...905...34H} have favored a singular or multiple dark and cool photospheric spots to explain the event\cite{2020ApJ...897L...9D} while preserving compatability with optical spectrophotometry, arguing in favor of a molecular opacity increase. A similar conclusion is reached by tomography from high resolution spectroscopy, which suggests that the propagation of two shock waves in the upper atmosphere, aided by underlying convection or outward gas motion altered the molecular opacity in the line of sight\cite{Kravchenko2020}. This scenario would be compatible with both anterior spectropolarimetric imaging\cite{2018A&A...620A.199L}, and 3D simulations of stellar convection in evolved stars carried out with the CO5BOLD code\cite{2002AN....323..213F,2012JCoPh.231..919F,2011A&A...535A..22C,2017A&A...600A.137F}. Here, convection is expected to trigger the formation of gas clumps capable of causing dimming events. In the present study, we explore both hypotheses: the formation of dust and photospheric cooling. 
	
	\noindent The cool patch hypothesis is explored by building a composite \textsc{Phoenix} model\cite{2007A&A...468..205L} of the stellar photosphere. The composite model images are then convolved with the ZIMPOL beam and compared to our observations. With the unperturbed photospheric temperature\cite{2020ApJ...891L..37L} set to 3,700~K, we determine the best cool patch temperatures to be 3,400~K in December 2019 and January 2020, down to 3,200~K in March 2020. The cool region has a best-fit extent of 62~\%, 79~\% and 43~\% of  the apparent disk, respectively, for the three epochs. All temperatures are given with a 50\,K uncertainty corresponding to the \textsc{Phoenix} grid spacing.
	
	\noindent The dust clump hypothesis is investigated through \textsc{Radmc3D}\cite{2012ascl.soft02015D} dust radiative transfer simulations. By exploring a grid of input parameters for a spherical dusty clump of constant density illuminated and heated by the RSG, we acquired optimized values for December 2019. For January and March 2020, the parameters were tuned manually to reproduce the peculiar shape of the images that proved very sensitive to the inputs. We retrieve clump radii of 4.5-6 astronomical unit (au), or $\sim$ 1~R$_\star$, and total dust masses of $(3-13) \times$ 10$^{-10}$~M$_\odot$. Considering the gas-to-dust ratio in the environment of RSG\cite{2011A&A...526A.156M} to be $\sim 200$, this would mean a total mass of the clump of $(0.7 - 3) \times$\,10$^{-7}$~M$_\odot$~(i.e., $2.3 - 8.5 \times$\,10$^{-2}$~M$_\mathrm{Earth}$). This represents 35 to 128~\% of the average annual mass loss of Betelgeuse for a low mass-loss rate estimation ($2.1 \times 10^{-7}$~M$_\odot$\,yr$^{-1}$, see [\cite{2010A&A...523A..18D}]), or 3 to 12~\% in the high range ($2 \pm 1 \times 10^{-6}$~M$_\odot$\,yr$^{-1}$, see [\cite{2016ApJ...819....7D}]). 
	
\begin{figure}[ht!]
	\centering
	\includegraphics[width=\columnwidth]{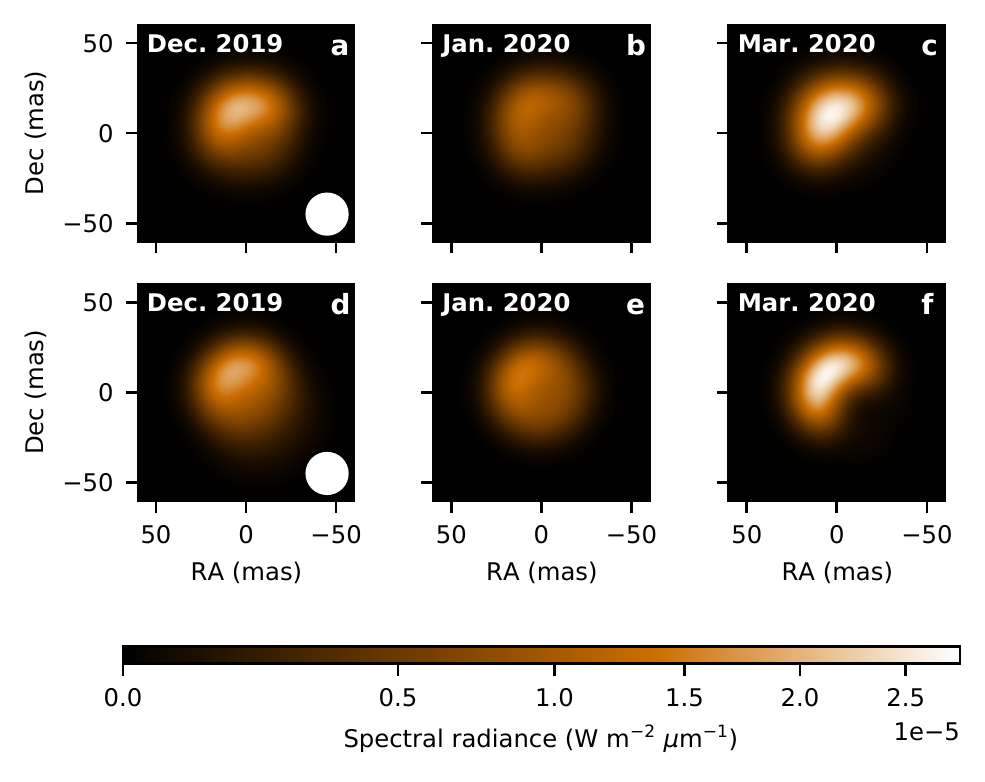}
	\caption{\textbf{Best model images obtained in the Cnt\_H$\alpha$ filter .} The images have been convolved with the SPHERE beam. \textbf{a}, \textbf{b}, and \textbf{c} correspond to the best matching \textsc{Phoenix} composite model (a cool spot) and \textbf{d}, \textbf{e}, and \textbf{f} to the best \textsc{Radmc3D} simulations (a dusty clump). We used a power-law scale intensity with index 0.65 to enhance the contrast.}
	\label{Fig:Img_Modeling_All_epochs}
\end{figure}

	\noindent The images of both types of models for the three dimmed observed epochs are plotted on Fig.~\ref{Fig:Img_Modeling_All_epochs}, showing a qualitative morphological agreement. Both the cool patch and dusty clump models capture the essential behavior during the Great Dimming, i.e. they both recover the level of optical dimming and the first-order atmospheric structure seen during the event. However, both models predict a J band near-infrared brightness decrease by a factor $1.2 - 1.3$ while a factor of just 1.02 is observed (Fig.~\ref{Fig:AAVSO_curve} and Extended Data Figure~\ref{Fig:photometry_SED}). Variations in the chemical composition, shape and properties of solid species in the dust occultation model, or a temporary change in photospheric molecular opacities, likely can reconcile the H-band flux and reproduce the American Association of Variable Star Observers (AAVSO) photometry in this spectral domain.\\
	
	
	
	\noindent Some observations of the Great Dimming clearly identify the presence of newly formed dust close to the photosphere\cite{2020RNAAS...4...39C,2020arXiv200505215S}, while other studies identify a cooling of the photosphere\cite{2020ApJ...897L...9D}. The Great Dimming occurred $424 \pm 4$ days after the star's previous minimum\cite{2020ATel13512....1G}. The primary pulsational period being $\sim 400$ days\cite{2010ApJ...725.1170S}, the scenario explaining the event must account for the alignment with the star's pulsational behavior. We propose that some time before the Great Dimming the star ejected a bubble of gas, likely at a favorable incidence in the pulsation cycle and potentially aided by the surfacing of a giant convective cell\cite{2020ApJ...899...68D}. The critical parameter to allow for dust condensation in the ambient environment of cool evolved stars is the temperature\cite{2019A&A...623A.158H}. The gas cloud may have been present in the near circumstellar environment but it would have been still too warm to trigger dust nucleation, until the star decreased in brightness in late 2019 in accordance with its pulsation phase. According to the \textsc{Radmc3D} modelling, with a local decrease of surface temperature from 3,700 to 3,400\,K in December 2019, the clump must have seen its environment temperature decrease from $2,300-1,900$ to $2,000-1,600$\,K at 12.5\,au which may have initiated a rapid dust formation\cite{2019MNRAS.489.4890B}. After the initial dust nucleation and obscuration, the gas further out is screened from the star, which can trigger a dust nucleation cascade causing the Great Dimming. Although a fully consistent \textit{ab-initio} 3D hydro-chemical model for such a scenario is not yet developed for an M-type RSG\cite{2019A&A...623A.158H}, its outline resembles remarkably the scenario proposed for R Coronae Borealis stars, including the coincidence of the dimming with the pulsational minimum\cite{1988MNRAS.233...65F}. \\
	
	
	\noindent Our observations from December 2019 through March 2020 for the first time disclose in real-time a discrete mass-loss event from a RSG star.  This mass loss is non-uniform and episodic, unambiguously linking the initiation mechanism to local surface behavior, i.e., a contrasted and rapidly changing photosphere. The released gas cloud experiences dust-nucleation within a few stellar radii, which may play an essential role in letting the ejected material escape from the system altogether.
	
	\noindent If such dusty-cloud ejections are a recurrent phenomenon, and observations of other RSGs suggest this may be the case\cite{2014A&A...568A..17O,2015A&A...584L..10S}, then only a fraction of these events may happen in the line of sight toward Earth and would lead to obscuration of stellar light. Betelgeuse's AAVSO light curve (Extended Data Figure~\ref{Fig:AAVSO_vis}) shows another possible dimming in 1984-1986, when some visual magnitude measurements dropped down to 2~mag. However, only two observers out of several tens saw it. And while these two visual observers saw Betelgeuse at a visual magnitude of 1.8-2, photoelectric measurements in the V band were at 0.7~mag. The 2019-2020 extreme dimming of Betelgeuse seems the only confirmed example for this star over the previous century.
	
	\noindent Previous observations of the circumstellar environment of Betelgeuse\cite{2009A&A...504..115K,2011A&A...531A.117K,2012AJ....144...36O,2012A&A...548A.113D,2018A&A...609A..67K} show a very inhomogeneous environment embedded in a smoother matrix. This may confirm that Betelgeuse and possibly other RSG experience two modes of mass loss\cite{2007AJ....133.2716H,2009AJ....137.3558S,2011A&A...531A.117K}: a smooth homogeneous radial outflow mainly consisting of gas with at most partial dust nucleation farther away from the star; and an episodic localised ejection of clumps of gas where conditions are favorable for efficient dust formation while still close to the photosphere. \\
	
	
	\noindent Photometric observations of the star have continued after the Great Dimming. Measurements\cite{2020ATel13901....1D} obtained in June-July 2020 showed that Betelgeuse experienced another dimming not corresponding to the 400 days period, but it has recovered since\cite{2020ATel13982....1S}. These successive dimmings fall within the irregular pattern of light curve variability of the star, and can be attributed to pulsation/convective activity\cite{2020ApJ...902...63J}.
	
	\noindent Our results confirm that the Great Dimming is not an indication of Betelgeuse's imminent explosion as a supernova\cite{2016ApJ...819....7D,2020ApJ...902...63J} (SN). Interpretation of the early behavior of core-collapse SNe light curves and post collapse spectral line variation point to enhanced pre-SN mass loss in the final weeks to centuries of the stars' life\cite{2017MNRAS.470.1642F}, in at least part of the progenitor population. For the longest timescale these enhanced rates may amount to $\sim 10^{-4}$\,M$_\odot$\,yr$^{-1}$; for the shorter timescales they may be as high as $\sim$1\,M$_\odot$\,yr$^{-1}$ [\cite{2017MNRAS.466.3021S}]. With a current mass-loss rate\cite{2010A&A...523A..18D,2016ApJ...819....7D} between $2 \times 10^{-7}$ and $2 \times 10^{-6}$\,M$_\odot$\,yr$^{-1}$\ Betelgeuse does not (yet) seem to have entered such a phase. Pre-SN activity possibly related to instabilities in nuclear burning\cite{2014ApJ...785...82S,2015ApJ...810...34W} or to waves driven by vigorous core convection\cite{2012MNRAS.423L..92Q,2017MNRAS.470.1642F} is reported for several Type IIP progenitors in the last few years before explosion\cite{2017NatPh..13..510Y,2010ApJ...715..541A}. A minority of the type IIP/L progenitor population may, however, show visual variability of no more than 5$-$10 percent in these final years, with the possible exception of short outbursts on the time scale of months\cite{2018MNRAS.480.1696J}. This means that some RSGs may show little or no sign of their impending core-collapse, years to weeks before it happens. Therefore, although the current mass loss behavior of Betelgeuse does not appear to forebode its demise, it remains possible that it may explode without prior warning.
	
	


\section*{Methods\label{Sect:Methods}}


\subsection*{VLT/SPHERE-ZIMPOL observations}

The resolved images were obtained through the Spectro-Polarimetric High-contrast Exoplanet REsearch (SPHERE\citeS{2019A&A...631A.155B}) instrument, mounted on the third unit telescope of ESO's Very Large Telescope (VLT). More precisely, we used one of its sub-systems: the Zurich IMaging POLarimeter (ZIMPOL\citeS{2014SPIE.9147E..3WR}). We observed Betelgeuse and a point spread function (PSF) calibrator, Rigel in the polarimetric P2 mode of ZIMPOL on December 31st 2018, December 26th 2019, January 27th 2020, March 18th and 20th 2020. With an angular diameter of $2.76 \pm 0.01$\,mas, Rigel\citeS{2010A&A...521A...5C} is well below the resolving power of ZIMPOL (24\,mas). Both stars were observed with several filters. The log of the observations is presented in Extended Data Table~\ref{Tab:Log_SPHERE}. We used the publicly available ESOreflex/SPHERE pipeline (v 0.38.0) to reduce the data and custom python routines\citeS{2015A&A...578A..77K} to produce the observables. This enabled us to derive the total intensity I, the polarized flux P, the degree of linear polarization DoLP, and the polarization electric-vector position angle $\theta$. The plate scale was $3.628\pm0.036$\,mas per pixel. The averaged beam size of the ZIMPOL observations was 24 milliarcseconds (mas) or 1.14 times Betelgeuse's radius. We performed a flux calibration of the ZIMPOL data using Rigel as a flux reference. However, the result showed discrepancies between the different filters and with the AAVSO measurements for the V filter (Extended Data Figure~\ref{Fig:photometry_SED}). We suspect that the issue results from the uncertainty on the transmission of the neutral densities used for the observations. Images at each epoch were deconvolved using the PyRAF\footnote{Available at \url{https://astroconda.readthedocs.io/en/latest/}} implemented Richardson-Lucy deconvolution algorithm, with Rigel as a measurement of the PSF. To avoid producing deconvolution artifacts, only 10 iterations were used on each frame. The result is visible on Fig.~\ref{Fig:ZIMPOL_SPHERE} and Extended Data Figure~\ref{Fig:Dec_ZIMPOL_Supp}.


\subsection*{VLT/SPHERE-IRDIS observations}

We used the sparse aperture masking (SAM) mode\citeS{2016SPIE.9907E..2TC} available on the Infrared Dual-band Imager and Spectrograph (IRDIS\citeS{2008SPIE.7014E..3LD}), another subsystem of SPHERE, to complement the GRAVITY observations. SAM uses a pupil mask with holes placed in a non-redundant configuration, designed to turn the single $8.2~\mathrm{m}$ dish telescope to an array of seven $1.2~\mathrm{m}$ diameter circular sub-apertures. Here the goal was to obtain interferometric measurements well within the first lobe of the visibility function to complement the VLTI/GRAVITY observations. Observations were taken in pupil tracking mode, with the SPHERE 7 holes masks, using two filters: NB\_CO and NB\_CntK2 ($\rm{\lambda_c}=$ 2,290 and  2,266\,nm respectively, with FWHM of 33 and 33\,nm). The log of the observations is presented in Extended Data Table~\ref{Tab:Log_SPHERE}. The data were reduced through the SPHERE data center\citeS{2017sf2a.conf..347D} applying the appropriate calibrations following the data reduction and handling pipeline\citeS{2008SPIE.7019E..39P} to correct for bad pixels, dark current, flat field, and sky background. Each frame is normalized in flux  and the corresponding parallactic angle is calculated. We then centered each image at the central lobe of the interferogram (using a 2D-Gaussian fit), cropped the images around this center, and manually sorted out the images of the cube to remove the bad frames with notable smearing or other source of error such as residual bad pixels. 

From the temporal cube of interferograms, the observables were extracted using a dedicated aperture masking data reduction software. The interferometric fringes created by each baseline of the non-redundant mask were fitted directly on the image plane \citeS{2011A&A...532A..72L, 2015ApJ...798...68G}. Our pipeline does the model fitting of the fringes using Single Value Decomposition (SVD), only the core of the interferogram is fitted, and the software includes a bandwidth smearing correction. From the amplitude and phase of the fringes squared visibilities and closure phases were obtained. Each frame in the data cubes were analyzed independently. Subsequently, the mean and standard deviation (s.d.) of the observables per data cube were computed. The mean observables were calibrated using the aperture masking observations of the point-like reference stars reported in Extended Data Table \ref{Tab:Log_SPHERE}. Finally, the different data sets per wavelengths were averaged into a unique OIFITS file that is used for the scientific analysis.  


\subsection*{VLTI/GRAVITY observations}

Betelgeuse and its interferometric calibrators were observed with GRAVITY\citeS{2017A&A...602A..94G} using the compact configuration (stations A0-B2-D0-C1) of the auxiliary telescopes (AT) of the VLT Interferometer (VLTI). We used the high spectral resolution ($\lambda/\Delta\lambda$ = 4,000) and dual polarization mode in order to accommodate for the brightness of Betelgeuse in the K band. The angular resolution reached was 14 mas. The log of the observations is given in Extended Data Table~\ref{Tab:Log_GRAVITY}. The data were reduced and calibrated through the ESOreflex pipeline in its version 1.2.4. We adopted the angular diameters of $2.242 \pm 0.212$\,mas for 56~Ori, and $3.364 \pm 0.283$\,mas for HD~44945 from the JMMC catalog\citeS{2016yCat.2345....0D}. After the initial calibration, the two polarizations were averaged.


\subsection*{Angular diameter determination}

The VLTI/GRAVITY and VLT/SPHERE-IRDIS SAM observations give us the squared visibility and closure phase as a function of the spatial frequency (sf = B/$\lambda$ with B the baseline length and $\lambda$ the wavelength). To estimate the angular diameter of the star, we built two datasets. The first one contains the data before the dimming (IRDIS of 2019-01-01 and GRAVITY of 2019-01-29), and the second one during the Great Dimming (IRDIS of 2019-12-27 and GRAVITY of 2020-02-14). As we do not expect the angular size of the star to change significantly on a scale of weeks, the time difference between the IRDIS and GRAVITY data of each set is negligible. We selected continuum data (CntK2 filter of the IRDIS observations and the range $2.22 - 2.28\,\mu$m for the GRAVITY data), thus excluding CO and water vapor absorption bands. Additionally we excluded some weak atomic and molecular absorption lines in the K band pseudo-continuum\citeS{2019A&A...621A...6O}. The angular diameter determination was done by fitting a uniform disk (UD) model to the squared visibility data only. This model seems initially justified by the limited deviation from a centro-symmetric model in the closure phase (Extended Data Figure~\ref{Fig:UD_vis2_fit_CP}) at low spatial frequency (for a centro-symmetric model, the closure phase should remain at 0$^\circ$ or 180$^\circ$). To avoid contamination by small scale structures, we limited the fit to the first lobe of visibility only (sf $< 6 \times 10^{6}$\,rad$^{-1}$). We estimated the angular diameter of Betelgeuse to be: before its dimming (January 2019) $\theta_\mathrm{UD} = 42.61 \pm 0.05$\,mas (reduced $\chi^2 = 26.5$), and during the dimming (December 2019 and February 2020) to $\theta_\mathrm{UD} = 42.11 \pm 0.05$\,mas (reduced $\chi^2 = 46.3$). Fitting a limb-darkened disk does not provide any further improvement and does not change significantly the angular diameter value. Therefore, since we are only interested in the angular diameter variation between the two epochs, we retain the UD fit reasonable for the first lobe of the data in the continuum.


\subsection*{Pre-existing dust extinction}

From previous studies\cite{2011A&A...531A.117K}\textsuperscript{,}\citeS{2009A&A...498..127V}, it is known that the circumstellar environment of Betelgeuse contains dust in an envelope around the star itself; this does not imply, however, that the dust is present in a homogeneous way. Before looking at the extinction caused by the Great Dimming, we need to assess the amount of dust that was already in the line of sight, whether from circumstellar or interstellar origin. Therefore, we took into account the AAVSO V, J and H band magnitude measurements obtained before the Dimming started, simultaneously with our January 2019 SPHERE observations. The AAVSO errorbars have been re-estimated from 0.01\,mag mostly to 0.1 mag to take into account the uncertainty on the calibrator star magnitudes. A possibility to model the already present circumstellar dust extinction is to include a spherical envelope around the star in a radiative transfer simulation. However, in order to limit the number of parameters, we decided instead to fit the A$_\mathrm{V}$ (extinction in the V band) that was required for the Cardelli\citeS{1989ApJ...345..245C} extinction law to reproduce a \textsc{Phoenix}\cite{2007A&A...468..205L} spectral energy distribution (SED) of a 15\,M$_\odot$~RSG. We adopted T$_\mathrm{eff} = 3,600$\,K (following spectrophotometric measurements\cite{2020ApJ...891L..37L}), and $\log g = 0.0$ (according to the literature\citeS{2019A&A...627A.138A} $\log g$ for Betelgeuse ranges from -0.32 to 0.43 with error bars up to 0.3). Since the ZIMPOL observations are performed through relatively broadband filters, we do not expect $\log g$ to be a sensitive parameter. Therefore, the dominant parameter in the selected SED is T$_\mathrm{eff}$, which is well constrained from the spectrophotometry. We adopted R$_\mathrm{V} = 4.2$ following RSG prescriptions\citeS{2005ApJ...634.1286M} (with R$_\mathrm{V} =$ A\textsubscript{V}/E\textsubscript{B-V}, E\textsubscript{B-V} being the color excess A\textsubscript{B} - A\textsubscript{V}). The result gives A$_\mathrm{V} = 0.65$. 


\subsection*{Local photospheric cooling hypothesis}

Considering the ZIMPOL images, if a dark cool spot is indeed at the origin of the Great Dimming, one cannot use a simple photospheric model to reproduce it since surfaces with different effective temperatures must coexist. To reproduce both our observations, we built a composite \textsc{Phoenix} model: we filled a stellar disk with a warm (normal photosphere) and a cool (anomalous patch) \textsc{Phoenix} photosphere\cite{2007A&A...468..205L}. The patch of cool temperature was circular with four parameters : its temperature, its radius and the coordinates (x, y) of its center. It is not allowed to expand outside the stellar disk. For each of the three epochs of the Dimming, we explored a grid of positions in the South-Western quadrant by steps of 0.5 mas. We used the same step size for the radius in the range 0 to 10 mas. We explored the temperature pairs (3,200; 3,700)K, (3,300; 3,700)K, and (3,400; 3,700)K for December 2019 and January 2020. For March 2020 we used the couples (3,200; 3700)K and (3,300; 3,800)K owing to the higher contrast between the bright and dark areas. Note that here the warm photosphere was set at 3,700\,K instead of the 3,600\,K measured from spectrophotometric measurements\cite{2020ApJ...891L..37L}, since we estimated that the later value represents an average on patches with various temperature. For each point of the grid, we computed the log-likelihood ($\mathcal{L}$) of the model image with respect to the ZIMPOL observations. The goal here was not to obtain a precise estimation of the T$_\mathrm{eff}$, but to assess the compatibility of such model with our images. Extended Data Figure~\ref{Fig:PHOENIX_all_Supp} shows the best images at each wavelength for each epoch. The best parameters are summarized in Extended Data Table~\ref{Tab:Result_Models}. Finally, the corresponding SEDs are plotted in Extended Data Figure~\ref{Fig:photometry_SED} together with the AAVSO and ZIMPOL photometry.


\subsection*{Dust clump hypothesis}

To test if the ZIMPOL images and AAVSO photometry can be reproduced by the presence of a dust clump in the line of sight to Betelgeuse, we used a simulation based on the radiative transfer code \textsc{Radmc3D}\cite{2012ascl.soft02015D}. The general scheme (coordinate system, physical parameters) of the simulation are illustrated in Extended Data Figure~\ref{Fig:RADMC3D_sketch}. We used a spherical grid of $20^3$ points sampling radii from 5 to 50\,au, longitudinal angles from 0 to $\pi$ and azimuthal angles from 0 to $2\pi$. We used five levels of adaptive mesh refinement to properly resolve the clump. For clarity, all the coordinates below will be given in the coordinate system described in Extended Data Figure~\ref{Fig:RADMC3D_sketch}. The star was modeled as a \textsc{Phoenix} photosphere\cite{2007A&A...468..205L} for a 15~M$_\odot$ RSG of T$_\mathrm{eff} = 3,700$\,K (3,800\,K for March 2020), and $\log g = 0.0$ according to parameters derived from spectrophotometric measurements\cite{2020ApJ...891L..37L}. We adopted the stellar size corresponding to the angular diameter measured with the VLT/SPHERE-IRDIS SAM and VLT/GRAVITY observations in December 2019 and January 2020, taking into account the distance\cite{2017AJ....154...11H} to the star. The dust clump was modeled as a spherical dust density centered at the coordinates ($x_c$, $y_c$, $z_c$), whose radius was $r_c$. The dust density was constant, $\rho_0$. To converge on the best parameters we ran a grid of \textsc{Radmc3D} models exploring  the range of parameters for $x_c$ and $y_c$  (0 to -3 au offset, with 5 steps), $z_c$ (5 to 20 au offset, with 5 steps), $r_c$ (4 to 8 au offset, with 5 steps), and $\rho_0$ (10$^{-19}$ to 10$^{-17}$ g\,cm$^{-3}$, with 5 steps logarithmically spaced). We adopted a canonical silicate composition for the dust (MgFeSiO$_4$)\citeS{1994A&A...292..641J,1995A&A...300..503D}. The sublimation temperature of this species is at about 1,500\,K, implying that for solutions that place the center of the clump at about 11\,au part of the spherical clump volume that is closest to the star is devoid of this species. However, in at least the outer half of the dusty sphere silicates can exist already at a clump center distance of 11\,au. This differentiation of the chemical composition inside of the clump does not impact our results (other than the actual dust distribution inside of the dusty clump). The grain size distribution is centered at 0.21~$\mu$m (ranging from 0.18 to 0.24\,$\mu$m), chosen to have the maximum dimming effect in the visible range (among distributions centered at 0.1, 0.21, 0.5, and 1 $\mu$m, using comparisons with spectra and images). The dust opacity parameters were computed using RADMC3D's dedicated python module based on Mie theory. The shape of the dust particles is assumed to be spherical since we are not interested in reproducing the polarized signal, the Gaussian grain size distribution was smeared out by 5~\% to avoid resonant effects. We refined these 3,125 models after a first iteration by adding two points to each parameter, surrounding the initial optimization. This lead to a total of 6,250 models. The best model was found by computing the log-likelihood for each point of the grid with respect to the ZIMPOL images. For January and March 2020 such a grid proved inefficient in finding a good enough combination of the parameters. Both January 2020 and March 2020 models are not optimized and are only best guesses. The best parameters are summarized in Extended Data Table~\ref{Tab:Result_Models}, the best images in each ZIMPOL filters are presented in Extended Data Figure~\ref{Fig:RADMC3D_all_Supp} and the best SED is plotted on Extended Data Figure~\ref{Fig:photometry_SED}.



\subsection*{Data availability}

Raw data were generated at the \href{http://archive.eso.org/cms.html}{European Southern Observatory} under ESO programs 0102.D-0240(A), 0102.D-0240(D), 104.20UZ, and 104.20V6.004. Derived data supporting the findings of this study are available at CDS via anonymous ftp to \url{cdsarc.u-strasbg.fr} (130.79.128.5) or via \url{http://cdsarc.u-strasbg.fr/viz-bin/qcat?J/other/Nat} (for the VLT/SPHERE-ZIMPOL images), and at the \href{http://oidb.jmmc.fr/index.html}{OiDB} (for the VLTI/GRAVITY and VLT/SPHERE-IRDIS SAM observations)

\subsection*{Code availability}

The SPHERE and GRAVITY pipelines are available on the ESO website: \url{http://www.eso.org/sci/software/pipelines/index.html}. The \textsc{Radmc3D} code is publicly available online: \url{https://github.com/dullemond/radmc3d-2.0}.


\subsection*{Acknowledgments}

We acknowledge with thanks the variable star observations from the AAVSO International Database contributed by observers worldwide and used in this research.
This project has received funding from the European Union's Horizon 2020 research and innovation program under the Marie Sk\l{}odowska-Curie Grant agreement No. 665501 with the research Foundation Flanders (FWO) ([PEGASUS]$^2$ Marie Curie fellowship 12U2717N awarded to M.M.). 
EC acknowledges funding from the KU Leuven C1 grant MAESTRO C16/17/007.
LD and MM acknowledge support from the ERC consolidator grant 646758 AEROSOL.
SK acknowledges support from the ERC starting grant 639889 ImagePlanetFormDiscs.
The material is based upon work supported by NASA under award number 80GSFC17M0002.
We are grateful to the ESO staff for their fast response in accepting the DDT proposal and carrying out the observations.
The authors would like to thank Betelgeuse for having undergone this peculiar event more than 700 years ago in the appropriate solid angle. 
This work has made use of the SPHERE Data Center, jointly operated by OSUG/IPAG (Grenoble), PYTHEAS/LAM/CeSAM (Marseille), OCA/Lagrange (Nice) and Observatoire  de Paris/LESIA (Paris).
This research has made use of the Jean-Marie Mariotti Center \texttt{Aspro} and \texttt{SearchCal} services\footnote{Available at \url{http://www.jmmc.fr}}. 
We used the SIMBAD and VIZIER databases at the CDS, Strasbourg (France)\footnote{Available at \url{http://cdsweb.u-strasbg.fr/}}, and NASA's Astrophysics Data System Bibliographic Services. 
This research made use of GNU Parallel\citeS{tange_ole_2018_1146014}, IPython\citeS{PER-GRA:2007}, Numpy\citeS{5725236}, Matplotlib\citeS{Hunter:2007}, SciPy\citeS{2020SciPy-NMeth}, Pandas\citeS{reback2020pandas,mckinney-proc-scipy-2010}, Astropy\footnote{Available at \url{http://www.astropy.org/}}, a community-developed core Python package for Astronomy\citeS{2013A&A...558A..33A}, and  Uncertainties\footnote{Available at \url{http://pythonhosted.org/uncertainties/}}: a Python package for calculations with uncertainties.\\

\noindent\textbf{Author contributions}\\
MM wrote the observing proposals, prepared all the observations, reduced and calibrated the ZIMPOL and GRAVITY data, ran the \textsc{Phoenix} and \textsc{Radmc3D} simulations, made all the figures of the article and is the main contributor to the text. EC cross checked the \textsc{Radmc3D} modeling. EL, JSB, and FC reduced the SPHERE-IRDIS data. AdK and LD wrote the discussion and conclusion of the main paper. All co-authors substantially contributed to the discussion, writing and revisions of the article.\\

\noindent\textbf{Author information}\\
Correspondence to \href{miguel.montarges@observatoiredeparis.psl.eu}{M. Montargès}.\\

\noindent\textbf{Competing interest statement}\\
The authors declare no competing interests.


\section*{Extended data}


\renewcommand{\tablename}{Extended Data Table}

\begin{table}[ht!]
	\sffamily
	\centering
	\caption{\textbf{Log of the VLT/SPHERE observations.} ND is the neutral density DIT is the detector integration time and NDIT the number of acquisitions. The airmass and seeing are provided by the observatory at the start of the acquisition. The filter and ND characteristics are available online \url{https://www.eso.org/sci/facilities/paranal/instruments/sphere/inst/filters.html}.}
	\label{Tab:Log_SPHERE}

	\begin{tabular}{c c c c c c c c c c}
		\hline
		\noalign{\smallskip}
		\multicolumn{10}{c}{VLT/SPHERE-ZIMPOL log}\\
		\hline
		\noalign{\smallskip}
		Date & Time (UT) & Target & Filter 1 & Filter 2 & ND & DIT (s) & NDIT & Airmass & Seeing (")\\
		\hline
		\noalign{\smallskip}
		2019-01-01 & 01:58 & Betelgeuse & Cnt\_H$\alpha$ & B\_H$\alpha$ & ND\_1 & 2 & 6 & 1.367 & 0.67\\
		& 02:26 & Rigel & Cnt\_H$\alpha$ & B\_H$\alpha$ & ND\_1 & 1 & 10 & 1.065 & 0.56\\
		& 04:00 & Betelgeuse & V & V & ND\_2 & 1.1 & 6 & 1.179 & 0.54\\
		& 04:14 & Rigel & V & V & ND\_2 & 1.1 & 6 & 1.076 & 0.5\\
		& 04:30 & Betelgeuse & KI & Cnt\_820 & ND\_2 & 3 & 6 & 1.194 & 0.77\\
		& 04:48 & Betelgeuse & TiO\_717 & Cnt\_748 & ND\_2 & 6 & 6 & 1.213 & 0.72\\
		& 05:25 & Rigel & KI & Cnt\_820 & ND\_2 & 6 & 6 & 1.233 & 0.67\\
		& 05:39 & Rigel & TiO\_717 & Cnt\_748 & ND\_2 & 3 & 6 & 1.274 & 0.54\\
		2019-12-27 & 03:27 & Betelgeuse & B\_H$\alpha$ & Cnt\_H$\alpha$ & ND\_1 & 1 & 50 & 1.208 & 0.52 \\
		& 03:46 & Rigel & B\_H$\alpha$ & Cnt\_H$\alpha$ & ND\_2 & 1 & 20 & 1.044 & 0.43 \\
		2020-01-28 & 02:00 & Betelgeuse &  B\_H$\alpha$ & Cnt\_H$\alpha$ & ND\_1 & 5 & 20 & 1.18 & 0.83 \\
		& 02:15 & Betelgeuse & V & V & ND\_2 & 5 & 20 & 1.179 & 0.84\\
		& 02:45 & Rigel &  B\_H$\alpha$ & Cnt\_H$\alpha$ & ND\_2 & 4 & 16 & 1.098 & 0.55 \\
		& 03:04 & Rigel & V & V & ND\_4 & 4 & 16 & 1.132 & 0.64\\
		& 03:21 & Betelgeuse & KI & Cnt\_H$\alpha$ & ND\_2 & 5 & 8 & 1.241 & 0.69\\
		& 03:33 & Betelgeuse & TiO\_717 & Cnt\_748 & ND\_4 & 4 & 8 & 1.266 & 0.69\\
		& 03:45 & Rigel & KI & Cnt\_820 & ND\_2 & 6 & 8 & 1.242 & 0.59\\
		& 03:57 & Rigel & TiO\_717 & Cnt\_748 & ND\_2 & 3 & 8 & 1.283 & 0.58\\
		2020-03-18 & 23:49 & Betelgeuse & KI & Cnt\_820 & ND\_2 & 2 & 20 & 1.223 & 0.74\\
		& 23:39 & Betelgeuse & TiO\_717 & Cnt\_748 & ND\_2 & 2 & 16 & 1.239 & 0.86\\
		2020-03-19 & 00:16 & Rigel & KI & Cnt\_820 & ND\_2 & 4 & 12 & 1.213 & 0.78\\
		& 00:26 & Rigel & TiO\_717 & Cnt\_748 & ND\_2 & 2 & 12 & 1.248 & 0.62\\
		2020-03-21 & 00:10 & Rigel & B\_H$\alpha$ & Cnt\_H$\alpha$ & ND\_2 & 4 & 14 & 1.218 & 0.75 \\
		& 00:27 & Rigel & V & V & ND\_4 & 20 & 6 & 1.28 & 0.73\\
		& 00:53 & Betelgeuse &  B\_H$\alpha$ & Cnt\_H$\alpha$ & ND\_2 & 5 & 18 & 1.42 & 0.38 \\
		& 01:11 & Betelgeuse & V & V & ND\_2 & 4 & 22 & 1.509 & 0.43\\
		\hline
		
		\noalign{\smallskip}
		\multicolumn{10}{c}{VLT/SPHERE-IRDIS log}\\
		
		\hline
		\noalign{\smallskip}
		Date & Time (UT) & Target & \multicolumn{2}{c}{Filter} & ND & NDIT & DIT & Airmass & Seeing (") \\
		\hline
		\noalign{\smallskip}
		2019-01-01 & 00:39 & $\phi$02 Ori & \multicolumn{2}{c}{NB\_CntK2} & NO & 100 & 4.0 & 1.745 & 0.76\\
		& 00:49 & $\phi$02 Ori & \multicolumn{2}{c}{NB\_CO} & NO & 100 & 4.0 & 1.670 & 0.71\\
		& 01:15 & Betelgeuse & \multicolumn{2}{c}{NB\_CntK2} & ND\_3.5 & 20 & 4.0 & 1.575 & 0.53\\
		& 01:17 & Betelgeuse & \multicolumn{2}{c}{NB\_CO} & ND\_3.5 & 20 & 4.0 & 1.557 & 0.55\\
		& 01:33 & 56 Ori & \multicolumn{2}{c}{NB\_CntK2} & NO & 20 & 0.837 & 1.365 & 0.68\\
		& 01:35 & 56 Ori & \multicolumn{2}{c}{NB\_CO} & NO & 20 & 0.837 & 1.358 & 0.68\\
		2019-12-27 & 04:35 & 56 Ori & \multicolumn{2}{c}{NB\_CO} & NO & 100 &0.837 & 1.122 & 0.39\\
		& 04:39 & 56 Ori & \multicolumn{2}{c}{NB\_CntK2} & NO & 100& 0.837  & 1.124 & 0.34\\
		& 04:05 & Betelgeuse & \multicolumn{2}{c}{NB\_CO} & ND\_3.5 & 100 & 3.0 & 1.181 & 0.33\\
		& 04:13 & Betelgeuse & \multicolumn{2}{c}{NB\_CntK2} & ND\_3.5 & 100 & 3.0 & 1.179 & 0.38\\
		\hline
	\end{tabular}
\end{table}

\begin{table}[ht!]
	\sffamily
	\centering
	\caption{\textbf{Log of the VLTI/GRAVITY observations on the A0-B2-D0-C1 quadruplet.} All the observations were executed in the high spectral resolution and dual polarization modes. The airmass and seeing are provided by the observatory at the start of the acquisition.}
	\label{Tab:Log_GRAVITY}
	
	\begin{tabular}{c c c c c c c}
		\hline
		\noalign{\smallskip}
		Date & Time (UT) & Target & Airmass & Seeing (")\\
		\hline
		\noalign{\smallskip}
		2019-01-29 & 00:31 & 56 Ori & 1.214 & 0.74 \\
		& 00:56 & Betelgeuse & 1.239 & 0.68 \\
		& 01:24 & HD 44945 & 1.194 & 0.95 \\
		& 01:47 & Betelgeuse & 1.183 & 1.06 \\
		& 02:14 & 56 Ori & 1.118 & 0.98 \\
		& 03:14 & 56 Ori & 1.175 & 0.97 \\
		& 03:36 & Betelgeuse & 1.286 & 0.94 \\
		2020-02-14 & 00:27 & Betelgeuse & 1.195 & 0.66 \\
		& 00:58 & HD 44945 & 1.150 & 0.55 \\
		& 01:49 & Betelgeuse & 1.185 & 0.46 \\
		& 02:28 & 56 Ori & 1.204 & 0.58 \\
		\hline
	\end{tabular}
	
\end{table}

\begin{table}[ht!]
	\sffamily
	\centering
	\caption{\textbf{Modeling results}. Best matched parameters for the composite \textsc{Phoenix} and dust clump \textsc{Radmc3D} models. For the \textsc{Phoenix} models, only the fraction of the cool patch recovering the stellar disk is kept in the solution. Pixels outside the stellar disk are set to 0 before convolution with the ZIMPOL PSF.}
	\label{Tab:Result_Models}
	\begin{tabular}{c c c c}
		\hline
		\noalign{\smallskip}
		\multicolumn{4}{c}{\textsc{Phoenix} composite patch}\\
		\hline
		\noalign{\smallskip}
		Parameter & December 2019 & January 2020 & March 2020 \\
		\hline
		\noalign{\smallskip}
		x$_\mathrm{center}$ (mas) & -7.1 & -2.4 & -28.4 \\
		y$_\mathrm{center}$ (mas) & -14.2 & -2.4 & -35.6 \\ 
		radius (mas) & 23.7 & 19.0 & 45.0 \\
		T$_\mathrm{hot}$ (K) & 3,700 & 3,700 & 3,700 \\
		T$_\mathrm{cool}$ (K) & 3,400 & 3,400 & 3,200 \\
		$\log \mathcal{L}$ & $-8.8 \times$\tento{6}& $-5.5 \times$\tento{7} & $-4.0 \times$\tento{7}\\
		\hline
		
		\noalign{\smallskip}
		\multicolumn{4}{c}{\textsc{Radmc3D} dusty clump}\\
		
		\hline
		\noalign{\smallskip}
		Parameter & December 2019 & January 2020 & March 2020 \\
		\hline
		\noalign{\smallskip}
		x$_\mathrm{c}$ (au) & -1.9 & -0.8 &  -1.9 \\
		y$_\mathrm{c}$ (au) & -3.0 & -0.6 & -1.8 \\ 
		z$_\mathrm{c}$ (au) & 12.5 & 20.0 & 20.0 \\ 
		r$_\mathrm{c}^{\mathrm{in}}$ (au) & 6.5 & 5.0 & 4.5 \\ 
		$\rho_0^{\mathrm{in}}$ (g\,cm$^{-3}$) & 3.2 $\times$ \tento{-19} &  4.0 $\times$ \tento{-19} & 2.0 $\times$ \tento{-18}\\
		\hline
	\end{tabular}	
\end{table}

\newpage


\renewcommand{\figurename}{Extended Data Figure}
\setcounter{figure}{0}

\begin{figure}[ht!]
	\centering
	\includegraphics[width=\columnwidth]{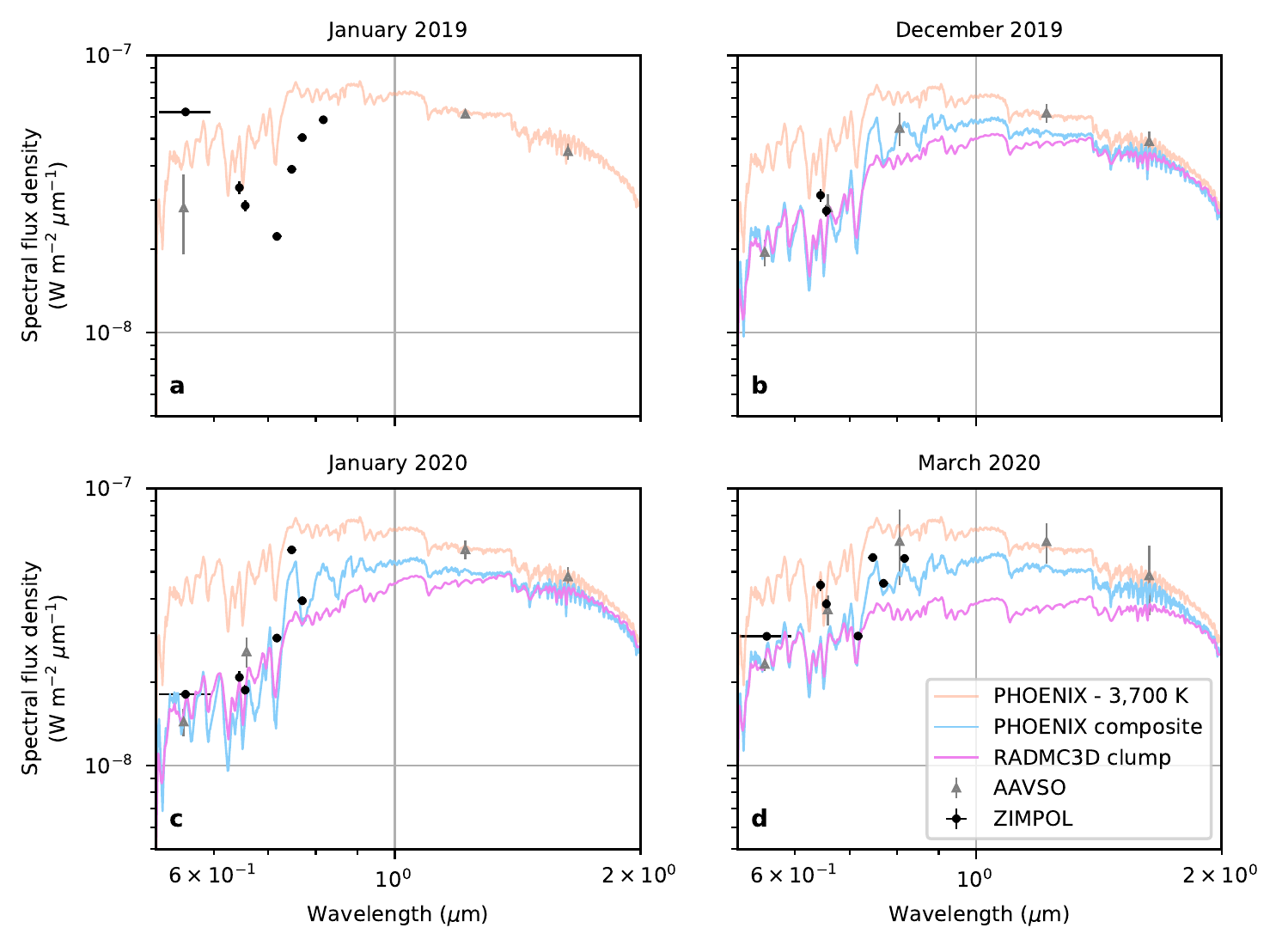}
	\caption{\textbf{Spectral energy distributions for the various epochs.} Photometry from the ZIMPOL filters (black dots) and from the AAVSO measurements (gray empty triangles) is compared to a 3,700~K \textsc{Phoenix} model (light orange), the best composite \textsc{Phoenix} model with a cool spot (blue) and the best \textsc{Radmc3D} dust clump model (violet). The flux errobars correspond to 1 s.d. The wavelength errorbars correspond to the width of the ZIMPOL filters. The AAVSO errorbars have been re-estimated from 0.01\,mag mostly to 0.1 mag to take into account the uncertainty on the calibrator star magnitudes.}
	\label{Fig:photometry_SED}
\end{figure}

\begin{figure}[ht!]
	\centering
	\includegraphics[width=\columnwidth]{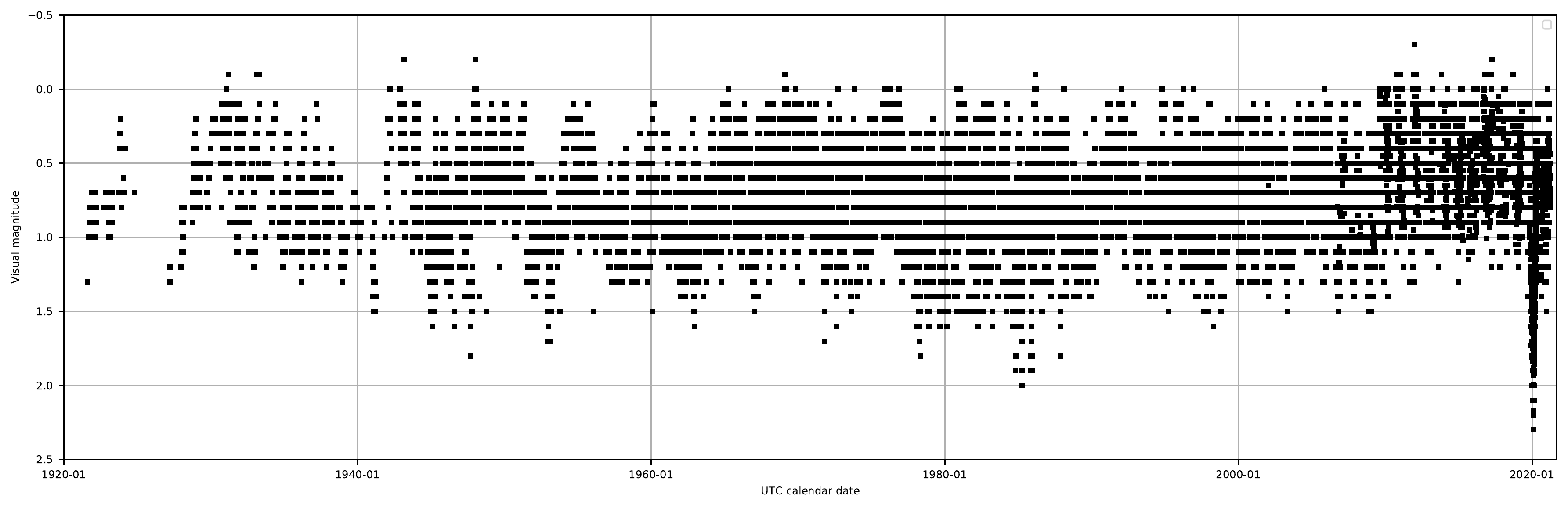}
	\caption{\textbf{Visual light curve of Betelgeuse.} The data are taken from the AAVSO database over the previous century.}
	\label{Fig:AAVSO_vis}
\end{figure}

\newpage

\begin{figure}[ht!]
	\centering
	\includegraphics[height=.8\textheight]{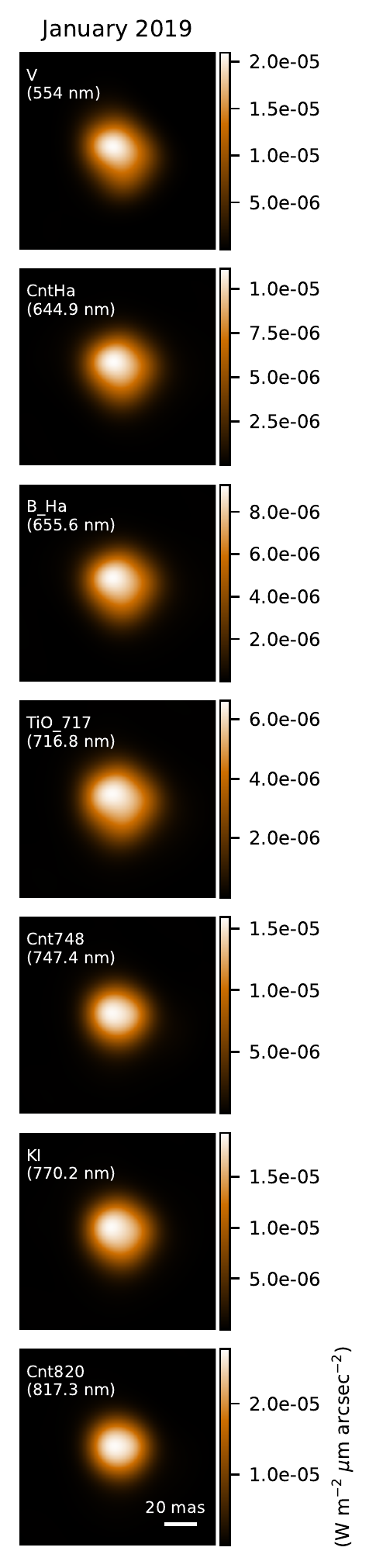}~
	\includegraphics[height=.8\textheight]{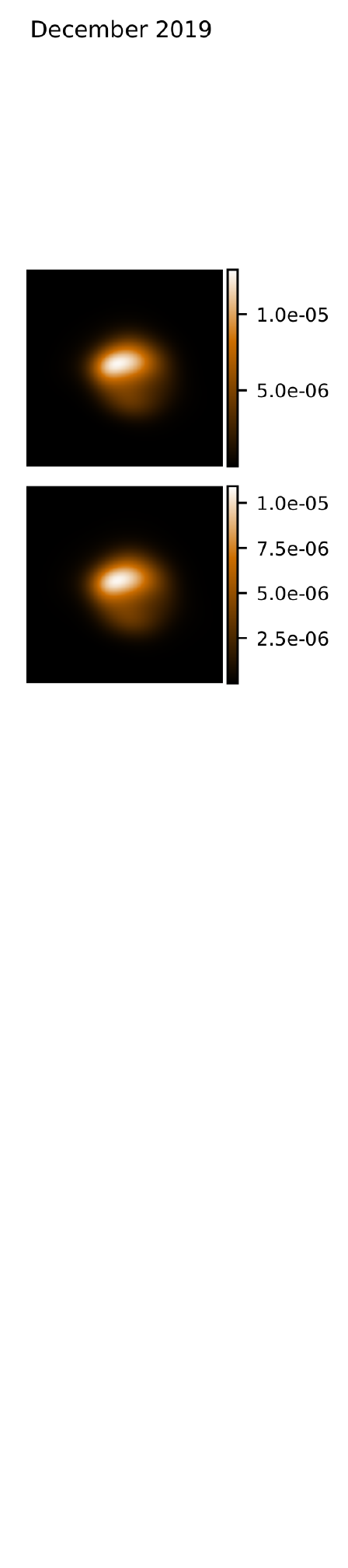}~
	\includegraphics[height=.8\textheight]{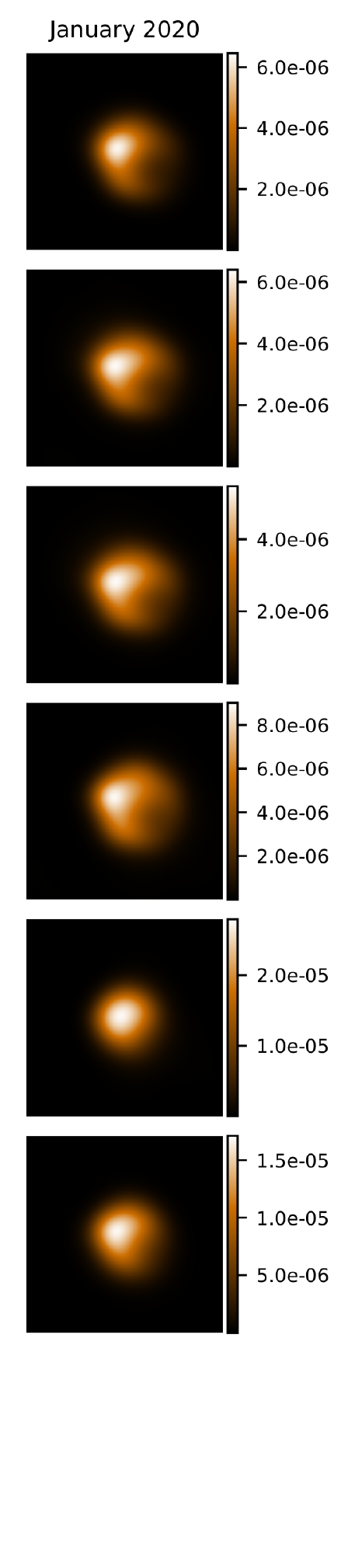}~
	\includegraphics[height=.8\textheight]{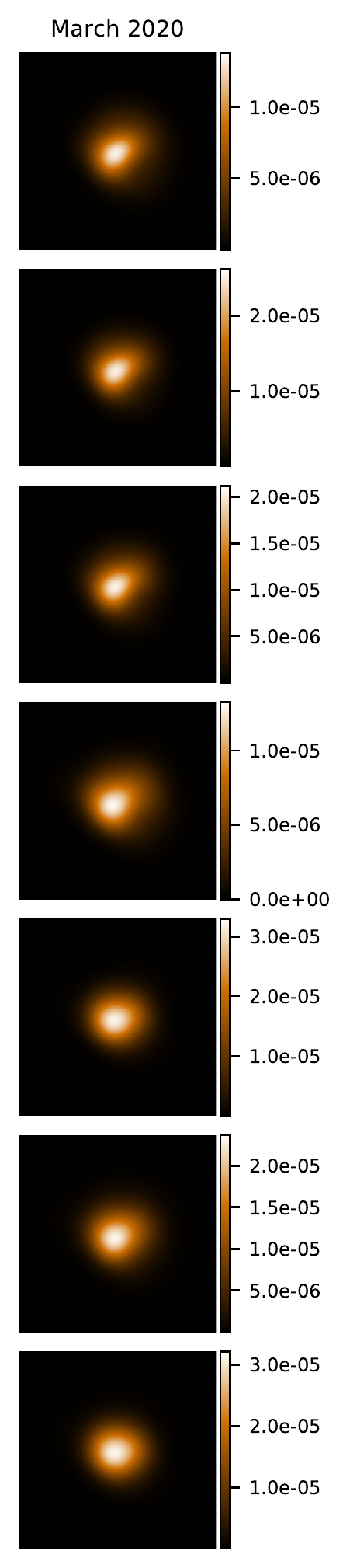}
	\caption{\textbf{Deconvolved intensity images of Betelgeuse for the various filters observed with ZIMPOL.} The spatial scale is indicated on the bottom left image. North is up, and East is left. Each row corresponds to a single filter. Each column corresponds to a single epoch. The colorscale is linear.}
	\label{Fig:Dec_ZIMPOL_Supp}
\end{figure}

\begin{figure}[ht!]
	\centering
	\includegraphics[width=.49\columnwidth]{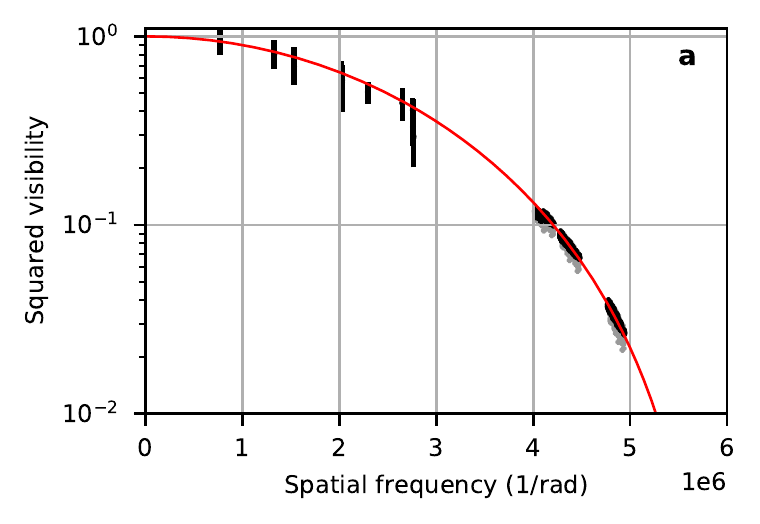}\,
	\includegraphics[width=.49\columnwidth]{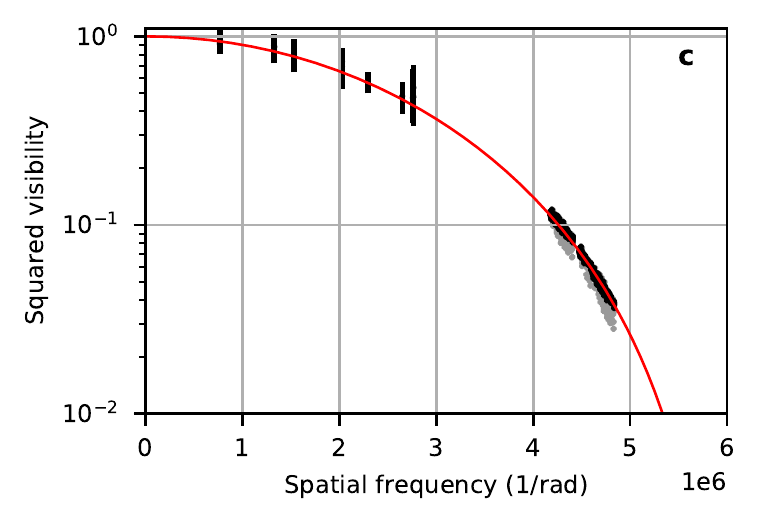}\\
	\includegraphics[width=.49\columnwidth]{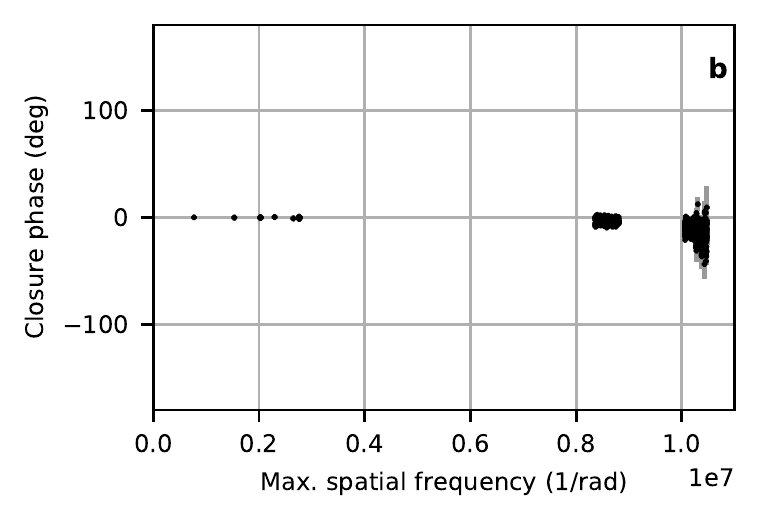}\,
	\includegraphics[width=.49\columnwidth]{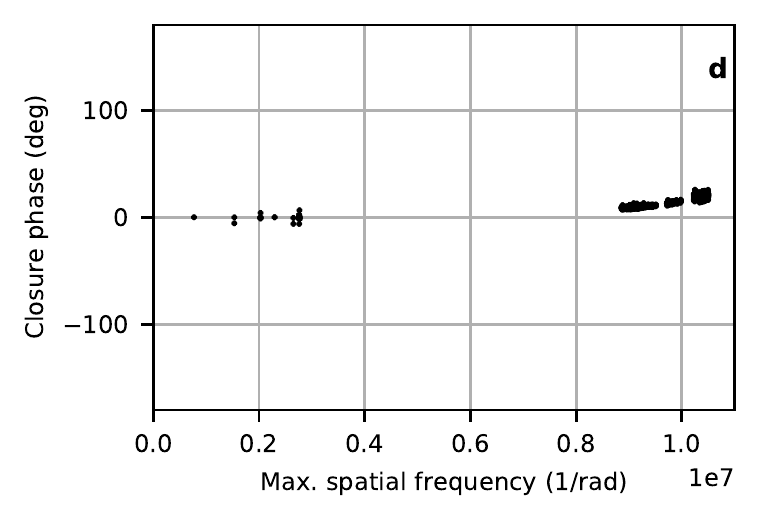}
	\caption{\textbf{Fit of the GRAVITY and IRDIS continuum data by a UD model.} The black points correspond to the data and the solid red curve to the model.  The gray points correspond to excluded photospheric lines. The errorbars correspond to 1 s.d. \textbf{a} shows the squared visibilities for January 2019, and \textbf{b} the corresponding closure phases. \textbf{c} shows the squared visibilities for February 2020, and \textbf{d} the corresponding closure phases.}
	\label{Fig:UD_vis2_fit_CP}
\end{figure}

\begin{figure}[ht!]
	\begin{center}
		\includegraphics[width=.3\columnwidth]{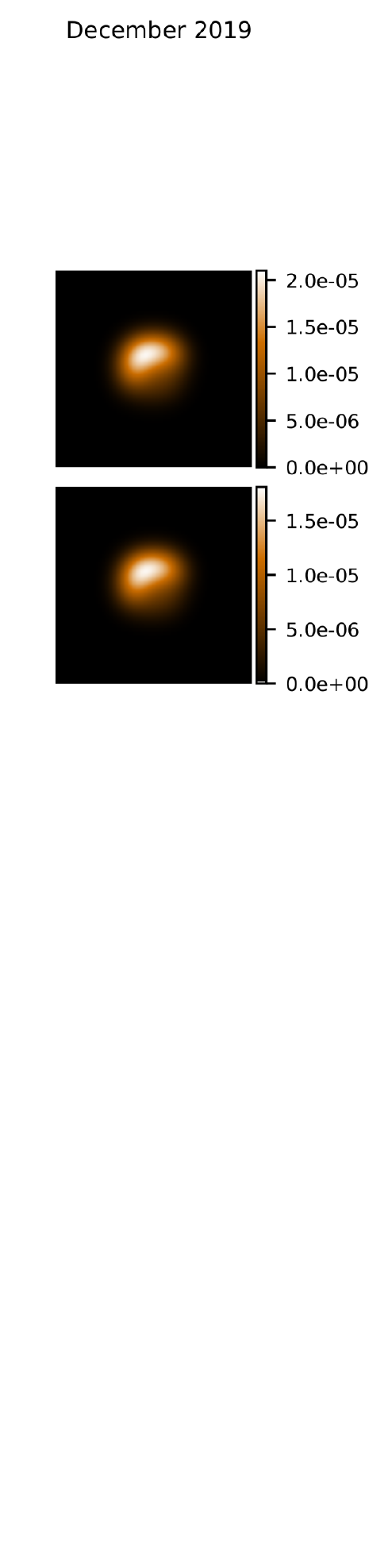}~
		\includegraphics[width=.3\columnwidth]{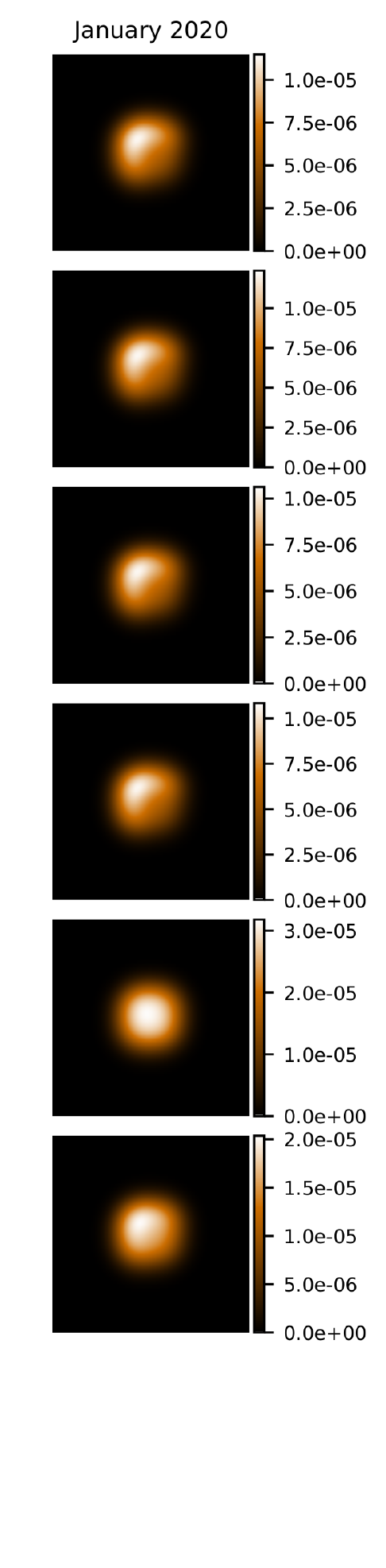}~
		\includegraphics[width=.3\columnwidth]{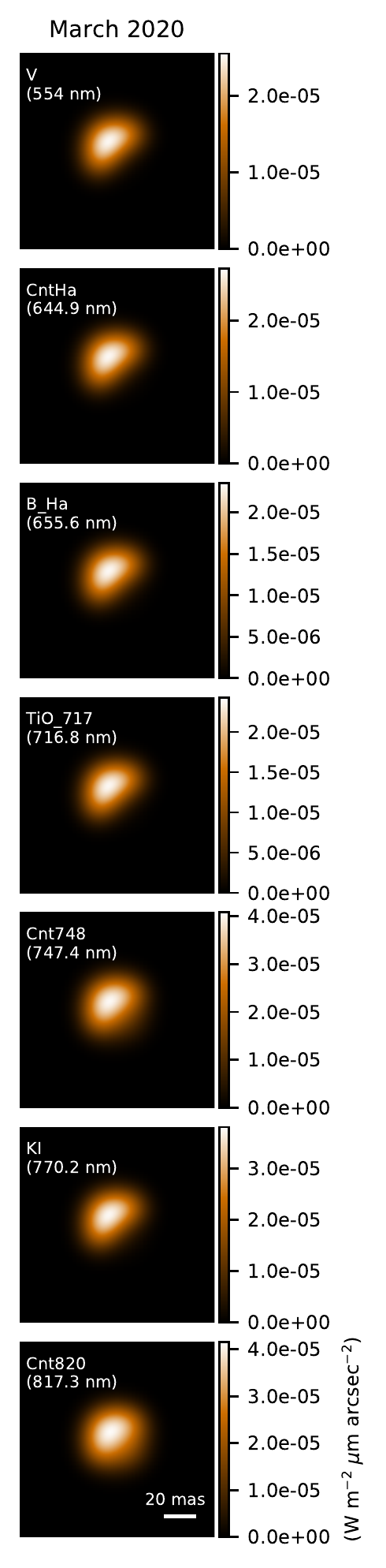}
	\end{center}
	\caption{\textbf{Best composite \textsc{Phoenix} model.} The spatial scale is indicated on the bottom right image. North is up, and East is left. Each row corresponds to a single filter. Each column corresponds to a single epoch. The colorscale is linear.\label{Fig:PHOENIX_all_Supp}}
\end{figure}

\begin{figure}[ht!]
	\centering
	\includegraphics[width=.7\columnwidth]{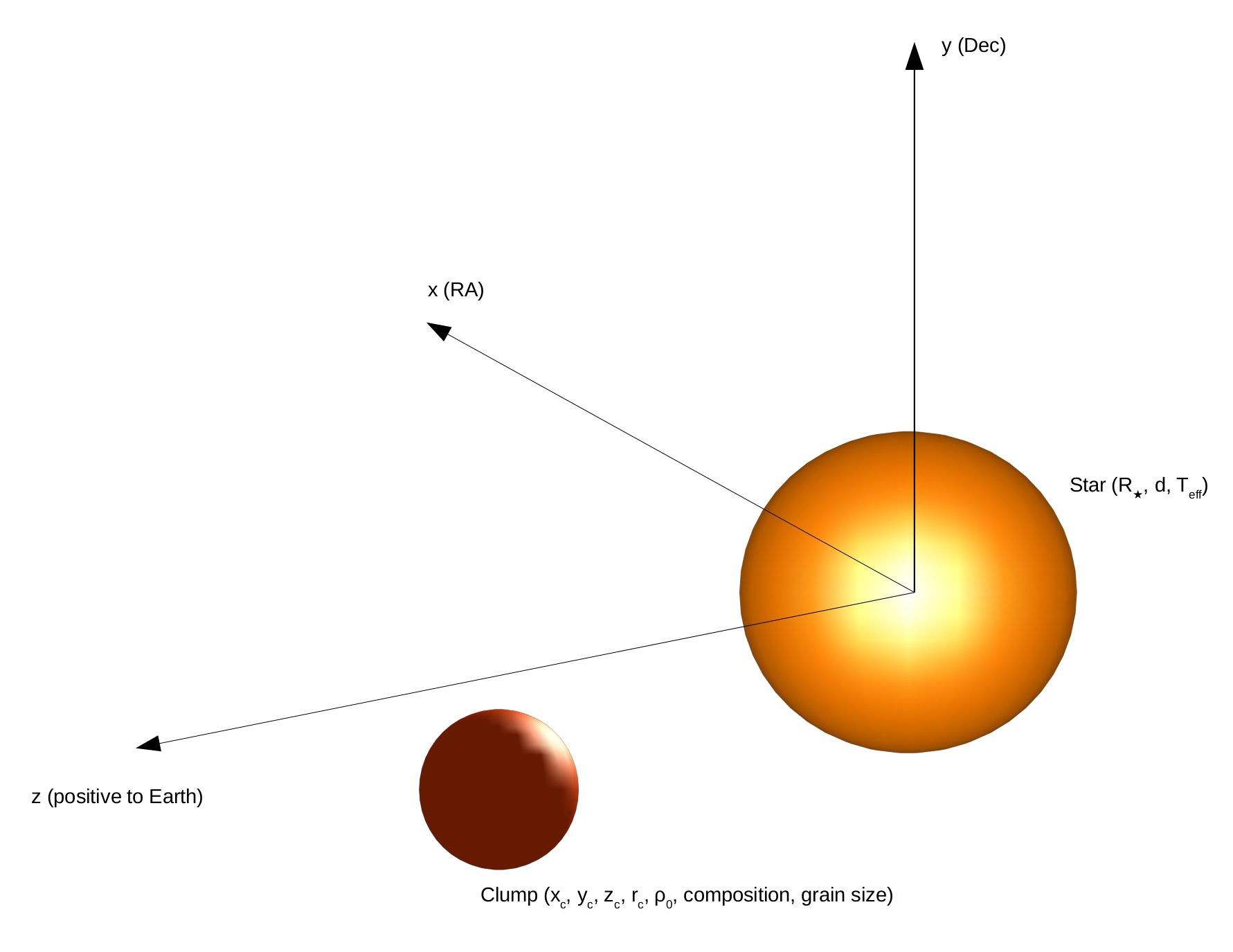}
	\caption{\textbf{Identification of the \textsc{Radmc3D} model.}}
	\label{Fig:RADMC3D_sketch}
\end{figure}

\begin{figure}[ht!]
	\begin{center}
		\includegraphics[width=.3\columnwidth]{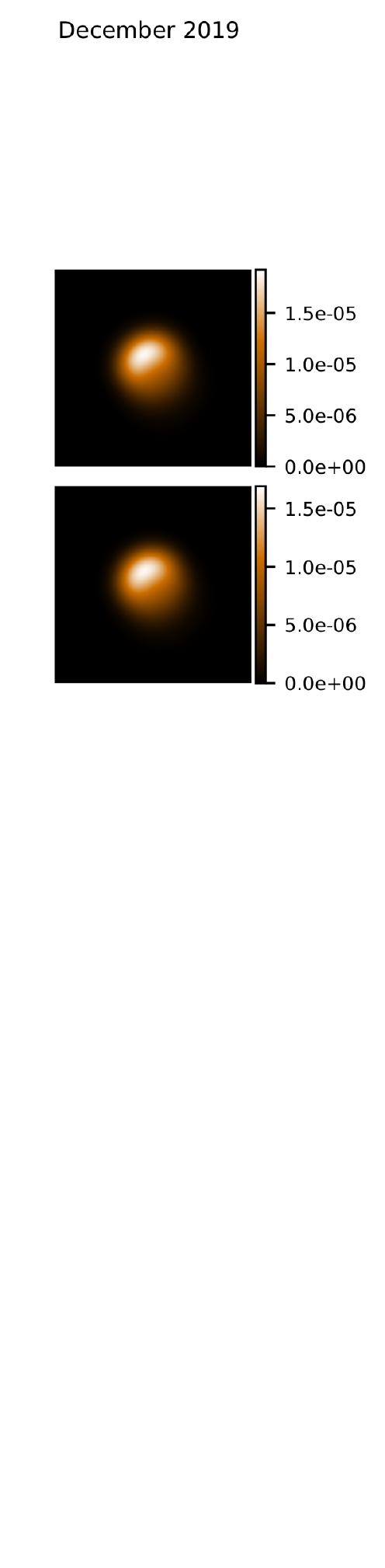}~
		\includegraphics[width=.3\columnwidth]{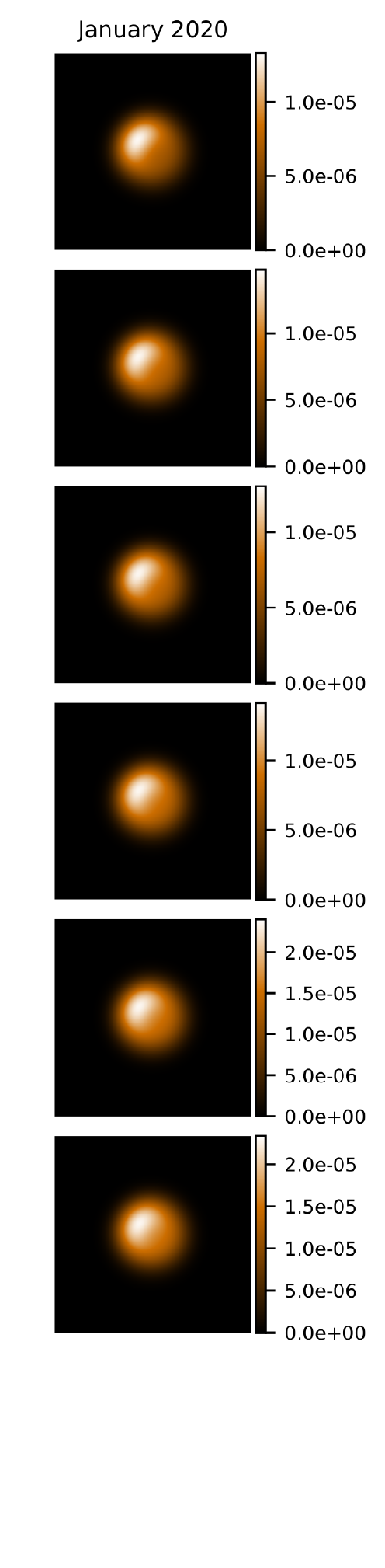}~
		\includegraphics[width=.3\columnwidth]{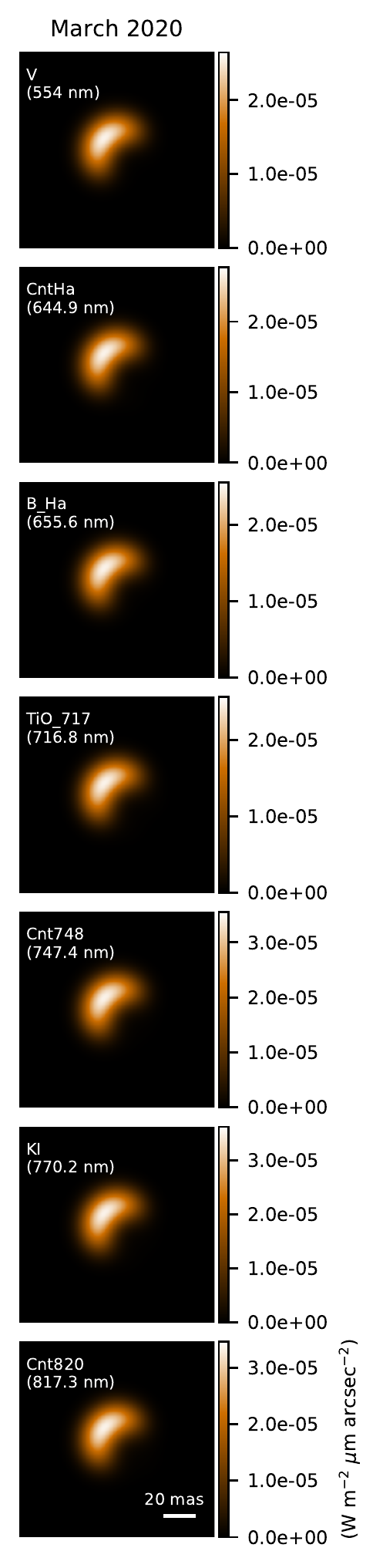}
	\end{center}
	\caption{\textbf{Best \textsc{Radmc3D} dusty clump models.} The spatial scale is indicated on the bottom right image. North is up, and East is left. Each row corresponds to a single filter.  Each column corresponds to a single epoch. The colorscale is linear.\label{Fig:RADMC3D_all_Supp}}
\end{figure}


\begin{thebibliography}{10}
		\urlstyle{rm}
		\expandafter\ifx\csname url\endcsname\relax
		\def\url#1{\texttt{#1}}\fi
		\expandafter\ifx\csname urlprefix\endcsname\relax\def\urlprefix{URL }\fi
		\expandafter\ifx\csname doiprefix\endcsname\relax\def\doiprefix{DOI: }\fi
		\providecommand{\bibinfo}[2]{#2}
		\providecommand{\eprint}[2][]{\url{#2}}
		
		\bibitem{2012A&A...537A.146E}
		\bibinfo{author}{{Ekstr{\"o}m}, S.} \emph{et~al.}
		\newblock \bibinfo{journal}{\bibinfo{title}{{Grids of stellar models with
					rotation. I. Models from 0.8 to 120 M$_{\&sun;}$ at solar metallicity (Z =
					0.014)}}}.
		\newblock {\emph{\JournalTitle{\aap}}} \textbf{\bibinfo{volume}{537}},
		\bibinfo{pages}{A146} (\bibinfo{year}{2012}).
		
		\bibitem{2015A&A...575A..50A}
		\bibinfo{author}{{Arroyo-Torres}, B.} \emph{et~al.}
		\newblock \bibinfo{journal}{\bibinfo{title}{{What causes the large extensions
					of red supergiant atmospheres?. Comparisons of interferometric observations
					with 1D hydrostatic, 3D convection, and 1D pulsating model atmospheres}}}.
		\newblock {\emph{\JournalTitle{\aap}}} \textbf{\bibinfo{volume}{575}},
		\bibinfo{pages}{A50} (\bibinfo{year}{2015}).
		
		\bibitem{2018MNRAS.476.2840M}
		\bibinfo{author}{Moriya, T.~J.}, \bibinfo{author}{F{\"o}rster, F.},
		\bibinfo{author}{Yoon, S.-C.}, \bibinfo{author}{Gr{\"a}fener, G.} \&
		\bibinfo{author}{Blinnikov, S.~I.}
		\newblock \bibinfo{journal}{\bibinfo{title}{{Type IIP supernova light curves
					affected by the acceleration of red supergiant winds}}}.
		\newblock {\emph{\JournalTitle{\mnras}}} \textbf{\bibinfo{volume}{476}},
		\bibinfo{pages}{2840--2851} (\bibinfo{year}{2018}).
		
		\bibitem{2015A&A...575A..60M}
		\bibinfo{author}{{Meynet}, G.} \emph{et~al.}
		\newblock \bibinfo{journal}{\bibinfo{title}{{Impact of mass-loss on the
					evolution and pre-supernova properties of red supergiants}}}.
		\newblock {\emph{\JournalTitle{\aap}}} \textbf{\bibinfo{volume}{575}},
		\bibinfo{pages}{A60} (\bibinfo{year}{2015}).
		
		\bibitem{2017AJ....154...11H}
		\bibinfo{author}{Harper, G.~M.} \emph{et~al.}
		\newblock \bibinfo{journal}{\bibinfo{title}{{An Updated 2017 Astrometric
					Solution for Betelgeuse}}}.
		\newblock {\emph{\JournalTitle{\aj}}} \textbf{\bibinfo{volume}{154}},
		\bibinfo{pages}{11} (\bibinfo{year}{2017}).
		
		\bibitem{2020ApJ...902...63J}
		\bibinfo{author}{{Joyce}, M.} \emph{et~al.}
		\newblock \bibinfo{journal}{\bibinfo{title}{{Standing on the Shoulders of
					Giants: New Mass and Distance Estimates for Betelgeuse through Combined
					Evolutionary, Asteroseismic, and Hydrodynamic Simulations with MESA}}}.
		\newblock {\emph{\JournalTitle{\apj}}} \textbf{\bibinfo{volume}{902}},
		\bibinfo{pages}{63} (\bibinfo{year}{2020}).
		
		\bibitem{2020ATel13512....1G}
		\bibinfo{author}{{Guinan}, E.}, \bibinfo{author}{{Wasatonic}, R.},
		\bibinfo{author}{{Calderwood}, T.} \& \bibinfo{author}{{Carona}, D.}
		\newblock \bibinfo{journal}{\bibinfo{title}{{The Fall and Rise in Brightness of
					Betelgeuse}}}.
		\newblock {\emph{\JournalTitle{The Astronomer's Telegram}}}
		\textbf{\bibinfo{volume}{13512}}, \bibinfo{pages}{1} (\bibinfo{year}{2020}).
		
		\bibitem{2011A&A...531A.117K}
		\bibinfo{author}{{Kervella}, P.} \emph{et~al.}
		\newblock \bibinfo{journal}{\bibinfo{title}{{The close circumstellar
					environment of Betelgeuse. II. Diffraction-limited spectro-imaging from 7.76
					to 19.50 {$\mu$}m with VLT/VISIR}}}.
		\newblock {\emph{\JournalTitle{\aap}}} \textbf{\bibinfo{volume}{531}},
		\bibinfo{pages}{A117} (\bibinfo{year}{2011}).
		
		\bibitem{2016A&A...585A..28K}
		\bibinfo{author}{{Kervella}, P.} \emph{et~al.}
		\newblock \bibinfo{journal}{\bibinfo{title}{{The close circumstellar
					environment of Betelgeuse. III. SPHERE/ZIMPOL imaging polarimetry in the
					visible}}}.
		\newblock {\emph{\JournalTitle{\aap}}} \textbf{\bibinfo{volume}{585}},
		\bibinfo{pages}{A28} (\bibinfo{year}{2016}).
		
		\bibitem{2011A&A...529A.163O}
		\bibinfo{author}{{Ohnaka}, K.} \emph{et~al.}
		\newblock \bibinfo{journal}{\bibinfo{title}{{Imaging the dynamical atmosphere
					of the red supergiant Betelgeuse in the CO first overtone lines with
					VLTI/AMBER}}}.
		\newblock {\emph{\JournalTitle{\aap}}} \textbf{\bibinfo{volume}{529}},
		\bibinfo{pages}{A163} (\bibinfo{year}{2011}).
		
		\bibitem{2020ApJ...891L..37L}
		\bibinfo{author}{{Levesque}, E.~M.} \& \bibinfo{author}{{Massey}, P.}
		\newblock \bibinfo{journal}{\bibinfo{title}{{Betelgeuse Just Is Not That Cool:
					Effective Temperature Alone Cannot Explain the Recent Dimming of
					Betelgeuse}}}.
		\newblock {\emph{\JournalTitle{\apjl}}} \textbf{\bibinfo{volume}{891}},
		\bibinfo{pages}{L37} (\bibinfo{year}{2020}).
		
		\bibitem{2020ApJ...905...34H}
		\bibinfo{author}{{Harper}, G.~M.}, \bibinfo{author}{{Guinan}, E.~F.},
		\bibinfo{author}{{Wasatonic}, R.} \& \bibinfo{author}{{Ryde}, N.}
		\newblock \bibinfo{journal}{\bibinfo{title}{{The Photospheric Temperatures of
					Betelgeuse during the Great Dimming of 2019/2020: No New Dust Required}}}.
		\newblock {\emph{\JournalTitle{\apj}}} \textbf{\bibinfo{volume}{905}},
		\bibinfo{pages}{34} (\bibinfo{year}{2020}).
		
		\bibitem{2020ApJ...897L...9D}
		\bibinfo{author}{{Dharmawardena}, T.~E.} \emph{et~al.}
		\newblock \bibinfo{journal}{\bibinfo{title}{{Betelgeuse Fainter in the
					Submillimeter Too: An Analysis of JCMT and APEX Monitoring during the Recent
					Optical Minimum}}}.
		\newblock {\emph{\JournalTitle{\apjl}}} \textbf{\bibinfo{volume}{897}},
		\bibinfo{pages}{L9} (\bibinfo{year}{2020}).
		
		\bibitem{Kravchenko2020}
		\bibinfo{author}{{Kravchenko}, K.} \emph{et~al.}
		\newblock \bibinfo{journal}{\bibinfo{title}{{Atmosphere of Betelgeuse before
					and during the great dimming revealed by tomography}}}.
		\newblock {\emph{\JournalTitle{\aap}}}  (\bibinfo{year}{subm.}).
		
		\bibitem{2018A&A...620A.199L}
		\bibinfo{author}{L{\'o}pez~Ariste, A.} \emph{et~al.}
		\newblock \bibinfo{journal}{\bibinfo{title}{{Convective cells in Betelgeuse:
					imaging through spectropolarimetry}}}.
		\newblock {\emph{\JournalTitle{\aap}}} \textbf{\bibinfo{volume}{620}},
		\bibinfo{pages}{A199} (\bibinfo{year}{2018}).
		
		\bibitem{2002AN....323..213F}
		\bibinfo{author}{{Freytag}, B.}, \bibinfo{author}{{Steffen}, M.} \&
		\bibinfo{author}{{Dorch}, B.}
		\newblock \bibinfo{journal}{\bibinfo{title}{{Spots on the surface of Betelgeuse
					-- Results from new 3D stellar convection models}}}.
		\newblock {\emph{\JournalTitle{Astronomische Nachrichten}}}
		\textbf{\bibinfo{volume}{323}}, \bibinfo{pages}{213--219}
		(\bibinfo{year}{2002}).
		
		\bibitem{2012JCoPh.231..919F}
		\bibinfo{author}{{Freytag}, B.} \emph{et~al.}
		\newblock \bibinfo{journal}{\bibinfo{title}{{Simulations of stellar convection
					with CO5BOLD}}}.
		\newblock {\emph{\JournalTitle{Journal of Computational Physics}}}
		\textbf{\bibinfo{volume}{231}}, \bibinfo{pages}{919--959}
		(\bibinfo{year}{2012}).
		
		\bibitem{2011A&A...535A..22C}
		\bibinfo{author}{{Chiavassa}, A.}, \bibinfo{author}{{Freytag}, B.},
		\bibinfo{author}{{Masseron}, T.} \& \bibinfo{author}{{Plez}, B.}
		\newblock \bibinfo{journal}{\bibinfo{title}{{Radiative hydrodynamics
					simulations of red supergiant stars. IV. Gray versus non-gray opacities}}}.
		\newblock {\emph{\JournalTitle{\aap}}} \textbf{\bibinfo{volume}{535}},
		\bibinfo{pages}{A22} (\bibinfo{year}{2011}).
		
		\bibitem{2017A&A...600A.137F}
		\bibinfo{author}{{Freytag}, B.}, \bibinfo{author}{{Liljegren}, S.} \&
		\bibinfo{author}{{H{\"o}fner}, S.}
		\newblock \bibinfo{journal}{\bibinfo{title}{{Global 3D radiation-hydrodynamics
					models of AGB stars. Effects of convection and radial pulsations on
					atmospheric structures}}}.
		\newblock {\emph{\JournalTitle{\aap}}} \textbf{\bibinfo{volume}{600}},
		\bibinfo{pages}{A137} (\bibinfo{year}{2017}).
		
		\bibitem{2007A&A...468..205L}
		\bibinfo{author}{{Lan{\c c}on}, A.}, \bibinfo{author}{{Hauschildt}, P.~H.},
		\bibinfo{author}{{Ladjal}, D.} \& \bibinfo{author}{{Mouhcine}, M.}
		\newblock \bibinfo{journal}{\bibinfo{title}{{Near-IR spectra of red supergiants
					and giants. I. Models with solar and with mixing-induced surface abundance
					ratios}}}.
		\newblock {\emph{\JournalTitle{\aap}}} \textbf{\bibinfo{volume}{468}},
		\bibinfo{pages}{205--220} (\bibinfo{year}{2007}).
		
		\bibitem{2012ascl.soft02015D}
		\bibinfo{author}{Dullemond, C.~P.} \emph{et~al.}
		\newblock \bibinfo{title}{{RADMC-3D: A multi-purpose radiative transfer tool}}
		(\bibinfo{year}{2012}).
		\newblock \eprint{1202.015}.
		
		\bibitem{2011A&A...526A.156M}
		\bibinfo{author}{{Mauron}, N.} \& \bibinfo{author}{{Josselin}, E.}
		\newblock \bibinfo{journal}{\bibinfo{title}{{The mass-loss rates of red
					supergiants and the de Jager prescription}}}.
		\newblock {\emph{\JournalTitle{\aap}}} \textbf{\bibinfo{volume}{526}},
		\bibinfo{pages}{A156} (\bibinfo{year}{2011}).
		
		\bibitem{2010A&A...523A..18D}
		\bibinfo{author}{{De Beck}, E.} \emph{et~al.}
		\newblock \bibinfo{journal}{\bibinfo{title}{{Probing the mass-loss history of
					AGB and red supergiant stars from CO rotational line profiles. II. CO line
					survey of evolved stars: derivation of mass-loss rate formulae}}}.
		\newblock {\emph{\JournalTitle{\aap}}} \textbf{\bibinfo{volume}{523}},
		\bibinfo{pages}{A18} (\bibinfo{year}{2010}).
		
		\bibitem{2016ApJ...819....7D}
		\bibinfo{author}{{Dolan}, M.~M.} \emph{et~al.}
		\newblock \bibinfo{journal}{\bibinfo{title}{{Evolutionary Tracks for
					Betelgeuse}}}.
		\newblock {\emph{\JournalTitle{\apj}}} \textbf{\bibinfo{volume}{819}},
		\bibinfo{pages}{7} (\bibinfo{year}{2016}).
		
		\bibitem{2020RNAAS...4...39C}
		\bibinfo{author}{{Cotton}, D.~V.}, \bibinfo{author}{{Bailey}, J.},
		\bibinfo{author}{{Horta}, A.~D.}, \bibinfo{author}{{Norris}, B. R.~M.} \&
		\bibinfo{author}{{Lomax}, J.~R.}
		\newblock \bibinfo{journal}{\bibinfo{title}{{Multi-band Aperture Polarimetry of
					Betelgeuse during the 2019-20 Dimming}}}.
		\newblock {\emph{\JournalTitle{Research Notes of the American Astronomical
					Society}}} \textbf{\bibinfo{volume}{4}}, \bibinfo{pages}{39}
		(\bibinfo{year}{2020}).
		
		\bibitem{2020arXiv200505215S}
		\bibinfo{author}{{Safonov}, B.} \emph{et~al.}
		\newblock \bibinfo{journal}{\bibinfo{title}{{Differential Speckle Polarimetry
					of Betelgeuse in 2019-2020: the rise is different from the fall}}}.
		\newblock {\emph{\JournalTitle{arXiv e-prints}}}
		\bibinfo{pages}{arXiv:2005.05215} (\bibinfo{year}{2020}).
		
		\bibitem{2010ApJ...725.1170S}
		\bibinfo{author}{{Stothers}, R.~B.}
		\newblock \bibinfo{journal}{\bibinfo{title}{{Giant Convection Cell Turnover as
					an Explanation of the Long Secondary Periods in Semiregular Red Variable
					Stars}}}.
		\newblock {\emph{\JournalTitle{\apj}}} \textbf{\bibinfo{volume}{725}},
		\bibinfo{pages}{1170--1174} (\bibinfo{year}{2010}).
		
		\bibitem{2020ApJ...899...68D}
		\bibinfo{author}{{Dupree}, A.~K.} \emph{et~al.}
		\newblock \bibinfo{journal}{\bibinfo{title}{{Spatially Resolved Ultraviolet
					Spectroscopy of the Great Dimming of Betelgeuse}}}.
		\newblock {\emph{\JournalTitle{\apj}}} \textbf{\bibinfo{volume}{899}},
		\bibinfo{pages}{68} (\bibinfo{year}{2020}).
		
		\bibitem{2019A&A...623A.158H}
		\bibinfo{author}{{H{\"o}fner}, S.} \& \bibinfo{author}{{Freytag}, B.}
		\newblock \bibinfo{journal}{\bibinfo{title}{{Exploring the origin of clumpy
					dust clouds around cool giants. A global 3D RHD model of a dust-forming
					M-type AGB star}}}.
		\newblock {\emph{\JournalTitle{\aap}}} \textbf{\bibinfo{volume}{623}},
		\bibinfo{pages}{A158} (\bibinfo{year}{2019}).
		
		\bibitem{2019MNRAS.489.4890B}
		\bibinfo{author}{{Boulangier}, J.}, \bibinfo{author}{{Gobrecht}, D.},
		\bibinfo{author}{{Decin}, L.}, \bibinfo{author}{{de Koter}, A.} \&
		\bibinfo{author}{{Yates}, J.}
		\newblock \bibinfo{journal}{\bibinfo{title}{{Developing a self-consistent AGB
					wind model - II. Non-classical, non-equilibrium polymer nucleation in a
					chemical mixture}}}.
		\newblock {\emph{\JournalTitle{\mnras}}} \textbf{\bibinfo{volume}{489}},
		\bibinfo{pages}{4890--4911} (\bibinfo{year}{2019}).
		
		\bibitem{1988MNRAS.233...65F}
		\bibinfo{author}{{Fadeev}, I.~A.}
		\newblock \bibinfo{journal}{\bibinfo{title}{{Carbon dust formation in R Coronae
					Borealis stars.}}}
		\newblock {\emph{\JournalTitle{\mnras}}} \textbf{\bibinfo{volume}{233}},
		\bibinfo{pages}{65--78} (\bibinfo{year}{1988}).
		
		\bibitem{2014A&A...568A..17O}
		\bibinfo{author}{{Ohnaka}, K.}
		\newblock \bibinfo{journal}{\bibinfo{title}{{Imaging the outward motions of
					clumpy dust clouds around the red supergiant Antares with VLT/VISIR}}}.
		\newblock {\emph{\JournalTitle{\aap}}} \textbf{\bibinfo{volume}{568}},
		\bibinfo{pages}{A17} (\bibinfo{year}{2014}).
		
		\bibitem{2015A&A...584L..10S}
		\bibinfo{author}{Scicluna, P.} \emph{et~al.}
		\newblock \bibinfo{journal}{\bibinfo{title}{{Large dust grains in the wind of
					VY Canis Majoris}}}.
		\newblock {\emph{\JournalTitle{\aap}}} \textbf{\bibinfo{volume}{584}},
		\bibinfo{pages}{L10} (\bibinfo{year}{2015}).
		
		\bibitem{2009A&A...504..115K}
		\bibinfo{author}{{Kervella}, P.} \emph{et~al.}
		\newblock \bibinfo{journal}{\bibinfo{title}{{The close circumstellar
					environment of Betelgeuse. Adaptive optics spectro-imaging in the near-IR
					with VLT/NACO}}}.
		\newblock {\emph{\JournalTitle{\aap}}} \textbf{\bibinfo{volume}{504}},
		\bibinfo{pages}{115--125} (\bibinfo{year}{2009}).
		
		\bibitem{2012AJ....144...36O}
		\bibinfo{author}{{O'Gorman}, E.} \emph{et~al.}
		\newblock \bibinfo{journal}{\bibinfo{title}{{CARMA CO(J = 2 - 1) Observations
					of the Circumstellar Envelope of Betelgeuse}}}.
		\newblock {\emph{\JournalTitle{\aj}}} \textbf{\bibinfo{volume}{144}},
		\bibinfo{pages}{36} (\bibinfo{year}{2012}).
		
		\bibitem{2012A&A...548A.113D}
		\bibinfo{author}{{Decin}, L.} \emph{et~al.}
		\newblock \bibinfo{journal}{\bibinfo{title}{{The enigmatic nature of the
					circumstellar envelope and bow shock surrounding Betelgeuse as revealed by
					Herschel. I. Evidence of clumps, multiple arcs, and a linear bar-like
					structure}}}.
		\newblock {\emph{\JournalTitle{\aap}}} \textbf{\bibinfo{volume}{548}},
		\bibinfo{pages}{A113} (\bibinfo{year}{2012}).
		
		\bibitem{2018A&A...609A..67K}
		\bibinfo{author}{Kervella, P.} \emph{et~al.}
		\newblock \bibinfo{journal}{\bibinfo{title}{{The close circumstellar
					environment of Betelgeuse. V. Rotation velocity and molecular envelope
					properties from ALMA}}}.
		\newblock {\emph{\JournalTitle{\aap}}} \textbf{\bibinfo{volume}{609}},
		\bibinfo{pages}{A67} (\bibinfo{year}{2018}).
		
		\bibitem{2007AJ....133.2716H}
		\bibinfo{author}{{Humphreys}, R.~M.}, \bibinfo{author}{{Helton}, L.~A.} \&
		\bibinfo{author}{{Jones}, T.~J.}
		\newblock \bibinfo{journal}{\bibinfo{title}{{The Three-Dimensional Morphology
					of VY Canis Majoris. I. The Kinematics of the Ejecta}}}.
		\newblock {\emph{\JournalTitle{\aj}}} \textbf{\bibinfo{volume}{133}},
		\bibinfo{pages}{2716--2729} (\bibinfo{year}{2007}).
		
		\bibitem{2009AJ....137.3558S}
		\bibinfo{author}{{Smith}, N.}, \bibinfo{author}{{Hinkle}, K.~H.} \&
		\bibinfo{author}{{Ryde}, N.}
		\newblock \bibinfo{journal}{\bibinfo{title}{{Red Supergiants as Potential Type
					IIn Supernova Progenitors: Spatially Resolved 4.6 {\ensuremath{\mu}}m CO
					Emission Around VY CMa and Betelgeuse}}}.
		\newblock {\emph{\JournalTitle{\aj}}} \textbf{\bibinfo{volume}{137}},
		\bibinfo{pages}{3558--3573} (\bibinfo{year}{2009}).
		
		\bibitem{2020ATel13901....1D}
		\bibinfo{author}{{Dupree}, A.}, \bibinfo{author}{{Guinan}, E.},
		\bibinfo{author}{{Thompson}, W.~T.} \& \bibinfo{author}{{STEREO/SECCHI/HI
				Consortium}}.
		\newblock \bibinfo{journal}{\bibinfo{title}{{Photometry of Betelgeuse with the
					STEREO Mission While in the Glare of the Sun from Earth}}}.
		\newblock {\emph{\JournalTitle{The Astronomer's Telegram}}}
		\textbf{\bibinfo{volume}{13901}}, \bibinfo{pages}{1} (\bibinfo{year}{2020}).
		
		\bibitem{2020ATel13982....1S}
		\bibinfo{author}{{Sigismondi}, C.} \emph{et~al.}
		\newblock \bibinfo{journal}{\bibinfo{title}{{Second dust cloud on
					Betelgeuse}}}.
		\newblock {\emph{\JournalTitle{The Astronomer's Telegram}}}
		\textbf{\bibinfo{volume}{13982}}, \bibinfo{pages}{1} (\bibinfo{year}{2020}).
		
		\bibitem{2017MNRAS.470.1642F}
		\bibinfo{author}{{Fuller}, J.}
		\newblock \bibinfo{journal}{\bibinfo{title}{{Pre-supernova outbursts via wave
					heating in massive stars - I. Red supergiants}}}.
		\newblock {\emph{\JournalTitle{\mnras}}} \textbf{\bibinfo{volume}{470}},
		\bibinfo{pages}{1642--1656} (\bibinfo{year}{2017}).
		
		\bibitem{2017MNRAS.466.3021S}
		\bibinfo{author}{{Smith}, N.} \emph{et~al.}
		\newblock \bibinfo{journal}{\bibinfo{title}{{Endurance of SN 2005ip after a
					decade: X-rays, radio and H{\ensuremath{\alpha}} like SN 1988Z require
					long-lived pre-supernova mass-loss}}}.
		\newblock {\emph{\JournalTitle{\mnras}}} \textbf{\bibinfo{volume}{466}},
		\bibinfo{pages}{3021--3034} (\bibinfo{year}{2017}).
		
		\bibitem{2014ApJ...785...82S}
		\bibinfo{author}{{Smith}, N.} \& \bibinfo{author}{{Arnett}, W.~D.}
		\newblock \bibinfo{journal}{\bibinfo{title}{{Preparing for an Explosion:
					Hydrodynamic Instabilities and Turbulence in Presupernovae}}}.
		\newblock {\emph{\JournalTitle{\apj}}} \textbf{\bibinfo{volume}{785}},
		\bibinfo{pages}{82} (\bibinfo{year}{2014}).
		
		\bibitem{2015ApJ...810...34W}
		\bibinfo{author}{{Woosley}, S.~E.} \& \bibinfo{author}{{Heger}, A.}
		\newblock \bibinfo{journal}{\bibinfo{title}{{The Remarkable Deaths of 9-11
					Solar Mass Stars}}}.
		\newblock {\emph{\JournalTitle{\apj}}} \textbf{\bibinfo{volume}{810}},
		\bibinfo{pages}{34} (\bibinfo{year}{2015}).
		
		\bibitem{2012MNRAS.423L..92Q}
		\bibinfo{author}{{Quataert}, E.} \& \bibinfo{author}{{Shiode}, J.}
		\newblock \bibinfo{journal}{\bibinfo{title}{{Wave-driven mass loss in the last
					year of stellar evolution: setting the stage for the most luminous
					core-collapse supernovae}}}.
		\newblock {\emph{\JournalTitle{\mnras}}} \textbf{\bibinfo{volume}{423}},
		\bibinfo{pages}{L92--L96} (\bibinfo{year}{2012}).
		
		\bibitem{2017NatPh..13..510Y}
		\bibinfo{author}{{Yaron}, O.} \emph{et~al.}
		\newblock \bibinfo{journal}{\bibinfo{title}{{Confined dense circumstellar
					material surrounding a regular type II supernova}}}.
		\newblock {\emph{\JournalTitle{Nature Physics}}} \textbf{\bibinfo{volume}{13}},
		\bibinfo{pages}{510--517} (\bibinfo{year}{2017}).
		
		\bibitem{2010ApJ...715..541A}
		\bibinfo{author}{{Andrews}, J.~E.} \emph{et~al.}
		\newblock \bibinfo{journal}{\bibinfo{title}{{SN 2007od: A Type IIP Supernova
					with Circumstellar Interaction}}}.
		\newblock {\emph{\JournalTitle{\apj}}} \textbf{\bibinfo{volume}{715}},
		\bibinfo{pages}{541--549} (\bibinfo{year}{2010}).
		
		\bibitem{2018MNRAS.480.1696J}
		\bibinfo{author}{{Johnson}, S.~A.}, \bibinfo{author}{{Kochanek}, C.~S.} \&
		\bibinfo{author}{{Adams}, S.~M.}
		\newblock \bibinfo{journal}{\bibinfo{title}{{The quiescent progenitors of four
					Type II-P/L supernovae}}}.
		\newblock {\emph{\JournalTitle{\mnras}}} \textbf{\bibinfo{volume}{480}},
		\bibinfo{pages}{1696--1704} (\bibinfo{year}{2018}).
		
	\end{thebibliography}

\begin{thebibliography}{20}
	\makeatletter
	\addtocounter{\@listctr}{49}
	\makeatother
	
	\urlstyle{rm}
	\expandafter\ifx\csname url\endcsname\relax
	\def\url#1{\texttt{#1}}\fi
	\expandafter\ifx\csname urlprefix\endcsname\relax\def\urlprefix{URL }\fi
	\expandafter\ifx\csname doiprefix\endcsname\relax\def\doiprefix{DOI: }\fi
	\providecommand{\bibinfo}[2]{#2}
	\providecommand{\eprint}[2][]{\url{#2}}
	
	\bibitem{2019A&A...631A.155B}
	\bibinfo{author}{{Beuzit}, J.~L.} \emph{et~al.}
	\newblock \bibinfo{journal}{\bibinfo{title}{{SPHERE: the exoplanet imager for
				the Very Large Telescope}}}.
	\newblock {\emph{\JournalTitle{\aap}}} \textbf{\bibinfo{volume}{631}},
	\bibinfo{pages}{A155} (\bibinfo{year}{2019}).
	
	\bibitem{2014SPIE.9147E..3WR}
	\bibinfo{author}{{Roelfsema}, R.} \emph{et~al.}
	\newblock \emph{\bibinfo{title}{{The ZIMPOL high contrast imaging polarimeter
				for SPHERE: system test results}}}, vol. \bibinfo{volume}{9147} of
	\emph{\bibinfo{series}{Society of Photo-Optical Instrumentation Engineers
			(SPIE) Conference Series}}, \bibinfo{pages}{91473W} (\bibinfo{year}{2014}).
	
	\bibitem{2010A&A...521A...5C}
	\bibinfo{author}{{Chesneau}, O.} \emph{et~al.}
	\newblock \bibinfo{journal}{\bibinfo{title}{{Time, spatial, and spectral
				resolution of the H{\ensuremath{\alpha}} line-formation region of Deneb and
				Rigel with the VEGA/CHARA interferometer}}}.
	\newblock {\emph{\JournalTitle{\aap}}} \textbf{\bibinfo{volume}{521}},
	\bibinfo{pages}{A5} (\bibinfo{year}{2010}).
	
	\bibitem{2015A&A...578A..77K}
	\bibinfo{author}{{Kervella}, P.} \emph{et~al.}
	\newblock \bibinfo{journal}{\bibinfo{title}{{The dust disk and companion of the
				nearby AGB star L$_{2}$ Puppis. SPHERE/ZIMPOL polarimetric imaging at visible
				wavelengths}}}.
	\newblock {\emph{\JournalTitle{\aap}}} \textbf{\bibinfo{volume}{578}},
	\bibinfo{pages}{A77} (\bibinfo{year}{2015}).
	
	\bibitem{2016SPIE.9907E..2TC}
	\bibinfo{author}{{Cheetham}, A.~C.} \emph{et~al.}
	\newblock \bibinfo{title}{{Sparse aperture masking with SPHERE}}.
	\newblock In \emph{\bibinfo{booktitle}{Optical and Infrared Interferometry and
			Imaging V}}, vol. \bibinfo{volume}{9907} of \emph{\bibinfo{series}{Society of
			Photo-Optical Instrumentation Engineers (SPIE) Conference Series}},
	\bibinfo{pages}{99072T} (\bibinfo{year}{2016}).
	
	\bibitem{2008SPIE.7014E..3LD}
	\bibinfo{author}{{Dohlen}, K.} \emph{et~al.}
	\newblock \emph{\bibinfo{title}{{The infra-red dual imaging and spectrograph
				for SPHERE: design and performance}}}, vol. \bibinfo{volume}{7014} of
	\emph{\bibinfo{series}{Society of Photo-Optical Instrumentation Engineers
			(SPIE) Conference Series}}, \bibinfo{pages}{70143L} (\bibinfo{year}{2008}).
	
	\bibitem{2017sf2a.conf..347D}
	\bibinfo{author}{{Delorme}, P.} \emph{et~al.}
	\newblock \bibinfo{title}{{The SPHERE Data Center: a reference for high
			contrast imaging processing}}.
	\newblock In \bibinfo{editor}{{Reyl{\'e}}, C.} \emph{et~al.} (eds.)
	\emph{\bibinfo{booktitle}{SF2A-2017: Proceedings of the Annual meeting of the
			French Society of Astronomy and Astrophysics}}, \bibinfo{pages}{Di}
	(\bibinfo{year}{2017}).
	\newblock \eprint{1712.06948}.
	
	\bibitem{2008SPIE.7019E..39P}
	\bibinfo{author}{{Pavlov}, A.} \emph{et~al.}
	\newblock \emph{\bibinfo{title}{{SPHERE data reduction and handling system:
				overview, project status, and development}}}, vol. \bibinfo{volume}{7019} of
	\emph{\bibinfo{series}{Society of Photo-Optical Instrumentation Engineers
			(SPIE) Conference Series}}, \bibinfo{pages}{701939} (\bibinfo{year}{2008}).
	
	\bibitem{2011A&A...532A..72L}
	\bibinfo{author}{{Lacour}, S.} \emph{et~al.}
	\newblock \bibinfo{journal}{\bibinfo{title}{{Sparse aperture masking at the
				VLT. I. Faint companion detection limits for the two debris disk stars HD
				92945 and HD 141569}}}.
	\newblock {\emph{\JournalTitle{\aap}}} \textbf{\bibinfo{volume}{532}},
	\bibinfo{pages}{A72} (\bibinfo{year}{2011}).
	
	\bibitem{2015ApJ...798...68G}
	\bibinfo{author}{{Greenbaum}, A.~Z.}, \bibinfo{author}{{Pueyo}, L.},
	\bibinfo{author}{{Sivaramakrishnan}, A.} \& \bibinfo{author}{{Lacour}, S.}
	\newblock \bibinfo{journal}{\bibinfo{title}{{An Image-plane Algorithm for
				JWST's Non-redundant Aperture Mask Data}}}.
	\newblock {\emph{\JournalTitle{\apj}}} \textbf{\bibinfo{volume}{798}},
	\bibinfo{pages}{68} (\bibinfo{year}{2015}).
	
	\bibitem{2017A&A...602A..94G}
	\bibinfo{author}{{Gravity Collaboration}} \emph{et~al.}
	\newblock \bibinfo{journal}{\bibinfo{title}{{First light for GRAVITY: Phase
				referencing optical interferometry for the Very Large Telescope
				Interferometer}}}.
	\newblock {\emph{\JournalTitle{\aap}}} \textbf{\bibinfo{volume}{602}},
	\bibinfo{pages}{A94} (\bibinfo{year}{2017}).
	
	\bibitem{2016yCat.2345....0D}
	\bibinfo{author}{Duvert, G.}
	\newblock \bibinfo{journal}{\bibinfo{title}{{VizieR Online Data Catalog: JMDC :
				JMMC Measured Stellar Diameters Catalogue (Duvert, 2016)}}}.
	\newblock {\emph{\JournalTitle{VizieR Online Data Catalog}}}
	\bibinfo{pages}{II/345} (\bibinfo{year}{2016}).
	
	\bibitem{2019A&A...621A...6O}
	\bibinfo{author}{{Ohnaka}, K.}, \bibinfo{author}{{Hadjara}, M.} \&
	\bibinfo{author}{{Maluenda Berna}, M.~Y.~L.}
	\newblock \bibinfo{journal}{\bibinfo{title}{{Spatially resolving the atmosphere
				of the non-Mira-type AGB star SW Vir in near-infrared molecular and atomic
				lines with VLTI/AMBER}}}.
	\newblock {\emph{\JournalTitle{\aap}}} \textbf{\bibinfo{volume}{621}},
	\bibinfo{pages}{A6} (\bibinfo{year}{2019}).
	
	\bibitem{2009A&A...498..127V}
	\bibinfo{author}{{Verhoelst}, T.} \emph{et~al.}
	\newblock \bibinfo{journal}{\bibinfo{title}{{The dust condensation sequence in
				red supergiant stars}}}.
	\newblock {\emph{\JournalTitle{\aap}}} \textbf{\bibinfo{volume}{498}},
	\bibinfo{pages}{127--138} (\bibinfo{year}{2009}).
	
	\bibitem{1989ApJ...345..245C}
	\bibinfo{author}{{Cardelli}, J.~A.}, \bibinfo{author}{{Clayton}, G.~C.} \&
	\bibinfo{author}{{Mathis}, J.~S.}
	\newblock \bibinfo{journal}{\bibinfo{title}{{The Relationship between Infrared,
				Optical, and Ultraviolet Extinction}}}.
	\newblock {\emph{\JournalTitle{\apj}}} \textbf{\bibinfo{volume}{345}},
	\bibinfo{pages}{245} (\bibinfo{year}{1989}).
	
	\bibitem{2019A&A...627A.138A}
	\bibinfo{author}{{Arentsen}, A.} \emph{et~al.}
	\newblock \bibinfo{journal}{\bibinfo{title}{{Stellar atmospheric parameters for
				754 spectra from the X-shooter Spectral Library}}}.
	\newblock {\emph{\JournalTitle{\aap}}} \textbf{\bibinfo{volume}{627}},
	\bibinfo{pages}{A138} (\bibinfo{year}{2019}).
	
	\bibitem{2005ApJ...634.1286M}
	\bibinfo{author}{{Massey}, P.} \emph{et~al.}
	\newblock \bibinfo{journal}{\bibinfo{title}{{The Reddening of Red Supergiants:
				When Smoke Gets in Your Eyes}}}.
	\newblock {\emph{\JournalTitle{\apj}}} \textbf{\bibinfo{volume}{634}},
	\bibinfo{pages}{1286--1292} (\bibinfo{year}{2005}).
	
	\bibitem{1994A&A...292..641J}
	\bibinfo{author}{{Jaeger}, C.}, \bibinfo{author}{{Mutschke}, H.},
	\bibinfo{author}{{Begemann}, B.}, \bibinfo{author}{{Dorschner}, J.} \&
	\bibinfo{author}{{Henning}, T.}
	\newblock \bibinfo{journal}{\bibinfo{title}{{Steps toward interstellar silicate
				mineralogy. 1: Laboratory results of a silicate glass of mean cosmic
				composition}}}.
	\newblock {\emph{\JournalTitle{\aap}}} \textbf{\bibinfo{volume}{292}},
	\bibinfo{pages}{641--655} (\bibinfo{year}{1994}).
	
	\bibitem{1995A&A...300..503D}
	\bibinfo{author}{{Dorschner}, J.}, \bibinfo{author}{{Begemann}, B.},
	\bibinfo{author}{{Henning}, T.}, \bibinfo{author}{{Jaeger}, C.} \&
	\bibinfo{author}{{Mutschke}, H.}
	\newblock \bibinfo{journal}{\bibinfo{title}{{Steps toward interstellar silicate
				mineralogy. II. Study of Mg-Fe-silicate glasses of variable composition.}}}
	\newblock {\emph{\JournalTitle{\aap}}} \textbf{\bibinfo{volume}{300}},
	\bibinfo{pages}{503} (\bibinfo{year}{1995}).
	
	\bibitem{tange_ole_2018_1146014}
	\bibinfo{author}{Tange, O.}
	\newblock \emph{\bibinfo{title}{{GNU Parallel 2018}}} (\bibinfo{publisher}{Ole
		Tange}, \bibinfo{year}{2018}).
	
	\bibitem{PER-GRA:2007}
	\bibinfo{author}{P\'erez, F.} \& \bibinfo{author}{Granger, B.~E.}
	\newblock \bibinfo{journal}{\bibinfo{title}{{IPython: a System for Interactive
				Scientific Computing}}}.
	\newblock {\emph{\JournalTitle{Computing in Science and Engineering}}}
	\textbf{\bibinfo{volume}{9}}, \bibinfo{pages}{21--29} (\bibinfo{year}{2007}).
	
	\bibitem{5725236}
	\bibinfo{author}{{van der Walt}, S.}, \bibinfo{author}{{Colbert}, S.~C.} \&
	\bibinfo{author}{{Varoquaux}, G.}
	\newblock \bibinfo{journal}{\bibinfo{title}{{The NumPy Array: A Structure for
				Efficient Numerical Computation}}}.
	\newblock {\emph{\JournalTitle{Computing in Science Engineering}}}
	\textbf{\bibinfo{volume}{13}}, \bibinfo{pages}{22--30}
	(\bibinfo{year}{2011}).
	
	\bibitem{Hunter:2007}
	\bibinfo{author}{Hunter, J.~D.}
	\newblock \bibinfo{journal}{\bibinfo{title}{{Matplotlib: A 2D graphics
				environment}}}.
	\newblock {\emph{\JournalTitle{Computing In Science \& Engineering}}}
	\textbf{\bibinfo{volume}{9}}, \bibinfo{pages}{90--95} (\bibinfo{year}{2007}).
	
	\bibitem{2020SciPy-NMeth}
	\bibinfo{author}{{Virtanen}, P.} \emph{et~al.}
	\newblock \bibinfo{journal}{\bibinfo{title}{{SciPy 1.0: Fundamental Algorithms
				for Scientific Computing in Python}}}.
	\newblock {\emph{\JournalTitle{Nature Methods}}} \textbf{\bibinfo{volume}{17}},
	\bibinfo{pages}{261--272} (\bibinfo{year}{2020}).
	
	\bibitem{reback2020pandas}
	\bibinfo{author}{pandas~development team, T.}
	\newblock \bibinfo{title}{{pandas-dev/pandas: Pandas}} (\bibinfo{year}{2020}).
	
	\bibitem{mckinney-proc-scipy-2010}
	\bibinfo{author}{{W}es {M}c{K}inney}.
	\newblock \bibinfo{title}{{{D}ata {S}tructures for {S}tatistical {C}omputing in
			{P}ython}}.
	\newblock In \bibinfo{editor}{{S}t\'efan van~der {W}alt} \&
	\bibinfo{editor}{{J}arrod {M}illman} (eds.)
	\emph{\bibinfo{booktitle}{{P}roceedings of the 9th {P}ython in {S}cience
			{C}onference}}, \bibinfo{pages}{56 -- 61} (\bibinfo{year}{2010}).
	
	\bibitem{2013A&A...558A..33A}
	\bibinfo{author}{{Astropy Collaboration}} \emph{et~al.}
	\newblock \bibinfo{journal}{\bibinfo{title}{{Astropy: A community Python
				package for astronomy}}}.
	\newblock {\emph{\JournalTitle{\aap}}} \textbf{\bibinfo{volume}{558}},
	\bibinfo{pages}{A33} (\bibinfo{year}{2013}).
	
\end{thebibliography}
\end{document}